\documentclass[twocolumn]{aastex63}
\usepackage{natbib}
\providecommand{\e}[1]{\ensuremath{\times 10^{#1}}}	            % for scientific notation
\providecommand{\HII}{\ion{H}{2}}           		        	% for H II regions
\providecommand{\HI}{\ion{H}{1}}               		        	% for H I regions
\providecommand{\OIII}{[\ion{O}{3}]}                        	% for [O III] lines
\providecommand{\OII}{[\ion{O}{2}]}                        	    % for [O II] lines
\providecommand{\SIII}{[\ion{S}{3}]}                    	    % for [S III] lines
\providecommand{\SII}{[\ion{S}{2}]}                        	    % for [S II] lines
\providecommand{\NII}{[\ion{N}{2}]}                        	    % for [N II] lines
\providecommand{\NeIII}{[\ion{Ne}{3}]}                        	% for [Ne III] lines
\providecommand{\NI}{[\ion{N}{1}]}                        	    % for [N I] lines
\providecommand{\OI}{[\ion{O}{1}]}                        	    % for [O I] lines
\providecommand{\ArIII}{[\ion{Ar}{3}]}                        	% for [Ar III] lines
\providecommand{\ArIV}{[\ion{Ar}{4}]}                        	% for [Ar IV] lines
\providecommand{\HA}{H$\alpha$}		                			% for H-Alpha
\providecommand{\HB}{H$\beta$}	                				% for H-Beta
\providecommand{\HG}{H$\gamma$}		                    		% for H-Gamma
\providecommand{\HD}{H$\delta$}		                			% for H-Delta
\providecommand{\HE}{H$\epsilon$} 	                			% for H-Epsilon
\providecommand{\R}{$R_{23}$}		                			% for R_23
\providecommand{\Te}{$T_{e}$}                   				% for T_e
\providecommand{\Ne}{$n_{e}$}	                				% for n_e
\providecommand{\cHB}{c$_{\text{H}\beta}$}	            		% for c(HB)
\providecommand{\abun}{12+$\log$(O/H)}	            			% for 12+log(O/H)
\providecommand{\LZ}{$L$--$Z$}		                			% for luminosity-metallicity
\providecommand{\MZ}{$M_{*}$--$Z$}	                			% for mass-metallicity
\providecommand{\LZR}{$LZR$}                                    % for LZR
\providecommand{\MZR}{$MZR$}                                    % for MZR
\providecommand{\Alecxy}{AGC~198691}            				% for the galaxy's name
\received{10 September 2021}
\revised{2 December 2021}
\accepted{3 December 2021}%\today}
\shorttitle{Abundances of \HA\ Dots}
\shortauthors{Hirschauer et al.}
\graphicspath{{./}{figures/}}
\begin{document}

\title{\HA\ Dots:\ Direct-Method Metal Abundances of Low-Luminosity Star-Forming Systems}

\correspondingauthor{Alec S.\ Hirschauer}
\email{ahirschauer@stsci.edu}

\author[0000-0002-2954-8622]{Alec S.\ Hirschauer}
\affil{Space Telescope Science Institute, 3700 San Martin Drive, Baltimore, MD 21218, USA}
%\email{ahirschauer@stsci.edu}

\author[0000-0001-8483-603X]{John J.\ Salzer}
\affil{Department of Astronomy, Indiana University, 727 East Third Street, Bloomington, IN 47405, USA}

\author[0000-0001-5247-1371]{Nathalie Haurberg}
\affil{Physics Department, Knox College, 2 East South Street, Galesburg, IL 61401, USA}

%\author[0000-0003-3810-3323]{Jesse R.\ Feddersen}
%\affil{Department of Astronomy, Yale University, P.O. Box 208101, New Haven, CT 06520-8101, USA}
%\affil{Department of Astronomy, Indiana University, 727 E.\ 3$^{rd}$ St., Bloomington, IN 47405, USA}

\author[0000-0001-6842-2371]{Caryl Gronwall}
\affiliation{Department of Astronomy \& Astrophysics, Pennsylvania State University, University Park, PA 16802, USA}
% 525 Davey Lab,
\affiliation{Institute for Gravitation \& the Cosmos, Pennsylvania State University, University Park, PA 16802, USA}

\author[0000-0001-9165-8905]{Steven Janowiecki}
\affiliation{University of Texas at Austin, McDonald Observatory, TX 79734, USA}



\begin{abstract}

\noindent Utilizing low-luminosity star-forming systems discovered in the \HA\ Dots survey, we present spectroscopic observations undertaken using the KPNO 4m telescope for twenty-six sources.
With determinations of robust, ``direct"-method metal abundances, we examine the properties of these dwarf systems, exploring their utility in characterizing starburst galaxies at low luminosities and stellar masses.
We find that the \HA\ Dots survey provides an effective new avenue for identifying star-forming galaxies in these regimes.
In addition, we examine abundance characteristics and metallicity scaling relations with these sources, highlighting a flattening of both the luminosity-metallicity (\LZ) and stellar mass-metallicity (\MZ) relation slopes in these regimes as compared with those utilizing samples covering wider respective dynamic ranges.
These local, accessible analogues to the kinds of star-forming dwarfs common at high redshift will help shed light on the building blocks which assembled into the massive galaxies common today.
%\looseness = -2

\end{abstract}



\section{Introduction} % Section 1.
\label{sec:introduction}

\indent An understanding of the comprehensive star formation and chemical enrichment histories of the universe depends critically upon the significant cumulative contributions of dwarf galaxies.
At the low metallicities typical of these low-luminosity star-forming systems, robust spectral abundances are obtainable via so-called ``direct methods," which rely upon detection and measurement of weakly-emitting, temperature-sensitive auroral emission lines, but whose strengths diminish with increasing metallicity.
Significant effort has been expended in such pursuit, with directed studies focusing on small samples (e.g., \citealp{bib:Skillman1989, bib:Lee2004, bib:MelbourneSalzer2002, bib:Izotov2006, bib:Ekta2008, bib:Brown2008, bib:Papaderos2008, bib:Hirschauer2015, bib:Berg2016, bib:Yang2017, bib:James2017, bib:Guseva2017, bib:Hsyu2018, bib:Izotov2019, bib:Berg2019, bib:Kojima2020}).
Wide-scale studies of star-forming systems, however, are strongly weighted toward larger numbers of galaxies that are more luminous (e.g., \citealp{bib:Tremonti2004}), a manifestation of the Malmquist effect.
Attempts to comprehensively characterize star-forming galaxy behavior generally rely on such wide-scale samples, thus leading to under-representation at the important dwarf regime.
In order to better understand galaxy chemical evolution overall, we must construct a link between these wide-scale programs with robust, carefully-selected samples focusing on low luminosities.
\\
\indent Ongoing endeavors to locate and identify low-luminosity, dwarf star-forming galaxies in the local universe typically rely on directed searches utilizing objective prism and narrowband filter surveys, with varying levels of success (e.g., \citealp{bib:KunthSargent1983, bib:KunthOstlin2000, bib:Salzer2000, bib:Salzer2001, bib:Izotov2012}).
Inevitably, selection biases limit the acquisition of a representative sample of sources, and thus galaxy samples are deficient in systems at the lowest luminosities.
Galaxy scaling relations such as luminosity-metallicity (\LZ) and stellar mass-metallicity (\MZ) are therefore typically weighted toward higher luminosities and, subsequently, stellar masses, thus under-representing the lower-valued regimes (e.g., \citealp{bib:Skillman1989, bib:RicherMcCall1995, bib:Lee2004, bib:vanZeeHaynes2006, bib:vanZee2006, bib:Lee2006, bib:Izotov2011, bib:Berg2012, bib:Haurberg2013, bib:Haurberg2015, bib:Hirschauer2016, bib:Calabro2017, bib:Hirschauer2018, bib:Blanc2019, bib:Indahl2021}).
\\
\indent Metallicity relation studies isolating the low-luminosity and low-mass ends of the star-forming galaxy distribution produce shallower slopes of these relation fits than for those made to wider-ranging samples.
These results suggest a fundamental difference in the chemical enrichment behavior experienced by star-forming dwarfs in these regimes \citep{bib:Blanc2019}.
An understanding of the heavy element enrichment characteristics in such systems, and how these directed studies are linked with larger-scale statistical samples of luminous galaxies, however, remains incomplete.
Abundance studies of metal-poor, low-luminosity galaxies are therefore of major importance for a full understanding of local star formation.
\\
\indent Study of these nearby, low-luminosity, low-abundance dwarfs can additionally provide insight regarding the behavior of star-forming systems at higher redshift, where detailed analyses are precluded due to the faintness of the sources.
These dwarf systems represent accessible analogues to the small galaxies thought to be ubiquitous at Cosmic Noon ($z$ $\sim$ 1.5--2; \citealp{bib:MadauDickinson2014}), which contributed significantly to the star formation and chemical enrichment of the universe.
Furthermore, such dwarf galaxies are expected to have coalesced to form the giant galaxies seen today (i.e., ``bottom-up"; \citealp{bib:WhiteRees1978, bib:FrenkWhite2012}).
A more complete picture of these systems will therefore inform our understanding of galaxy evolution from the Big Bang to the present.
\\
\indent The \HA\ Dots project \citep{bib:Kellar2012, bib:Salzer2020, bib:Watkins2021} has identified a new sample of low-luminosity, dwarf star-forming systems.
These sources were serendipitously found within images of the ALFALFA \HA\ survey (hereafter AHA; \citealp{bib:VanSistine2016}).
AHA observed galaxies of the Arecibo Legacy Fast ALFA project (ALFALFA; \citealp{bib:Giovanelli2005, bib:Haynes2011, bib:Haynes2018}), a large-scale blind \HI\ survey.
%, via emission through the narrowband \HA\ filter of the WIYN 0.9m telescope.
%\footnote{The 0.9m telescope is operated by WIYN Inc.\ on behalf of a Consortium of nine partner Universities and Organizations (see \url{https://www.noao.edu/0.9m/partners}).
%WIYN is a joint partnership of the University of Wisconsin at Madison, Indiana University, and the National Optical Astronomical Observatory (NOAO).}.
The origin of the \HA\ Dots flux can include \HA\ from local sources or shorter wavelength emission lines redshifted into the \HA\ filter transmission window.
Determination of the origins of this flux is determined via low-resolution ``quick-look" spectroscopy.
Generally point-like in appearance and not obviously associated with the original AHA target galaxy, those \HA\ Dots found at low redshift ($z$ = 0.0019 -- 0.0243; a recessional velocity range of approximately 500 -- 7500 km s$^{-1}$) represent some of the lowest-luminosity, isolated, dwarf star-forming systems known in the local universe.
\\
\indent In this paper, we present directed spectral observations for a set of twenty-six local, low-luminosity, \HA-emitting \HA\ Dot sources.
We present spectral line flux measurements and estimations of stellar mass via spectral energy distribution (SED) fitting techniques.
From these data we produce robust chemical abundances determined via the direct method, utilizing the temperature-sensitive \OIII$\lambda$4363 auroral line.
While thus far comprising only a small subset of the entire \HA\ Dots catalog, these low-luminosity and low-mass sources will be exceptionally valuable in improving our understanding of the full spectrum of starburst galaxies in the local universe.
In addition, detailed abundance analyses of these galaxies will provide an important observational constraint for models of chemical enrichment in the early universe (e.g., \citealp{bib:Greif2010}).
In \S \ref{sec:data} we describe the observational program carried out for \HA\ Dots and the data reduction and analysis processes.
Section \ref{sec:analysis} details the determination of chemical abundances, while \S \ref{sec:discussion} investigates the properties and characteristics of these objects.
We summarize our results in \S \ref{sec:summary}.
Throughout this work, we assume a standard cosmology of $\Omega_{\Lambda}$ = 0.73, $\Omega_{M}$ = 0.27, and $H_{0}$ = 70 km s$^{-1}$ Mpc$^{-1}$.
% from HA Dots II paper:
% 85 of 154 unique objects are H-alpha-selected (55%)
% redshift range 0.0056 - 0.0243 (1690 - 7430 km/s)
% first-look spectra obtained with HET LRS, but no [OII]3727 sensitivity
% NB: HA Dot 127 a compact, outlying HII region of a LSB dwarf SF galaxy
% Should I include an ELG type parameter from HA Dots II?
% e.g., BCD, outlying HII region, GP-like galaxy (<- only [OIII]-selected though)
% Potentially to replace "Distance [Mpc]" in Table 1, since it's redundant

\begin{deluxetable*}{ccccccccccccc} % Table 1
\rotate
\tablenum{1}
\label{tab:Dot_info}
\tabletypesize{\small}
\tablewidth{0pt}
\tablecaption{
Coordinate, redshift, distance, photometric, stellar mass, star-formation rate, specific star-formation rate, observing run, and spectroscopic instrument information for \HA\ Dots in this study.
}
\tablehead{
\colhead{\HA\ Dot}&\colhead{RA}&\colhead{Dec}&\colhead{Redshift}&\colhead{Distance}&\colhead{$m_{R}$}&\colhead{$M_{R}$}&\colhead{$M_{B}$}&\colhead{M$_{*}$}&\colhead{SFR}&\colhead{sSFR}&\colhead{Obs.\ Run}&\colhead{Instrument}
\\
\colhead{}&\colhead{[degrees]}&\colhead{[degrees]}&\colhead{[$z$]}&\colhead{[Mpc]}&\colhead{}&\colhead{}&\colhead{}&\colhead{[log $M_{\odot}$]}&\colhead{[$M_{\odot}\ yr^{-1}$]}&\colhead{[log SFR M$_{*}^{-1}$]}&\colhead{(Date)}&\colhead{(RC/K)}
}
\startdata
2 & 131.81389 & 10.04178 & 0.01093 & 44.9 & 18.61 & --14.77 & --14.26 & 7.23 & --1.51 & --8.73 & April 2012 & RC Spec. \\
--- & --- & --- & --- & --- & --- & --- & --- & --- & --- & --- & April 2015 & KOSMOS \\
4 & 131.82460 & 10.04264 & 0.01108 & 46.7 & 17.12 & --16.35 & --16.08 & 7.94 & --0.80 & --8.75 & April 2012 & RC Spec. \\
12 & 192.73056 & 12.02537 & 0.00605 & 24.5 & 18.06 & --13.98 & --13.19 & 7.07 & --2.28 & --9.35 & April 2012 & RC Spec. \\
20 & 39.70882 & 27.79560 & 0.01481 & 65.9 & 18.25 & --16.23 & --14.94 & 7.88 & --0.94 & --8.81 & October 2012 & RC Spec. \\
31 & 209.27116 & 14.08034 & 0.01486 & 65.0 & 19.07 & --15.04 & --14.66 & 7.39 & --1.40 & --8.78 & April 2012 & RC Spec. \\
34 & 217.68205 & 13.95239 & 0.01723 & 75.5 & 17.45 & --17.00 & --16.38 & 8.26 & --1.09 & --9.35 & April 2012 & RC Spec. \\
40 & 246.68385 & 11.62925 & 0.01654 & 72.0 & 18.64 & --15.77 & --15.32 & 7.88 & --1.64 & --9.51 & April 2012 & RC Spec. \\
43 & 349.67995 & 26.22570 & 0.02063 & 94.5 & 18.59 & --16.46 & --15.45 & 7.99 & --1.48 & --9.47 & September 2014 & KOSMOS \\
--- & --- & --- & --- & --- & --- & --- & --- & --- & --- & --- & December 2016 & KOSMOS \\
47 & 10.15819 & 27.04192 & 0.01696 & 76.4 & 21.83 & --12.71 & --12.04 & 6.03 & --1.64 & --7.67 & October 2012 & RC Spec. \\
53 & 12.43780 & 27.13195 & 0.02312 & 103.4 & 17.57 & --17.61 & --16.70 & 8.60 & --0.41 & --9.01 & September 2014 & KOSMOS \\
79 & 11.15322 & 26.92934 & 0.01709 & 76.0 & 18.43 & --16.07 & --15.44 & 7.68 & --1.03 & --8.71 & September 2014 & KOSMOS \\
81 & 2.03670 & 27.45397 & 0.01512 & 69.8 & 18.85 & --15.50 & --15.07 & 7.56 & --0.98 & --8.54 & October 2012 & RC Spec. \\
--- & --- & --- & --- & --- & --- & --- & --- & --- & --- & --- & September 2014 & KOSMOS \\
90 & 15.86986 & 26.80917 & 0.01682 & 76.0 & 17.70 & --16.89 & --16.18 & 8.07 & --0.19 & --8.26 & September 2014 & KOSMOS \\
116 & 211.20357 & 11.69860 & 0.01568 & 68.1 & 16.68 & --17.53 & --17.04 & 8.56 & --0.67 & --9.23 & April 2012 & RC Spec. \\
124 & 351.35497 & 25.19517 & 0.01670 & 76.0 & 17.34 & --17.19 & --16.33 & 8.44 & --0.79 & --9.22 & September 2014 & KOSMOS \\
127 & 5.99423 & 25.17786 & 0.01491 & 67.2 & 18.24 & --15.97 & --13.92 & 7.92 & --1.61 & --9.53 & December 2016 & KOSMOS \\
131 & 29.64904 & 24.65515 & 0.01666 & 75.1 & 18.69 & --15.90 & --15.13 & 7.90 & --1.07 & --8.97 & September 2014 & KOSMOS \\
138 & 34.00316 & 25.22497 & 0.01689 & 75.5 & 18.80 & --15.76 & --14.76 & 7.74 & --1.43 & --9.18 & September 2014 & KOSMOS \\
145 & 136.62712 &  5.18636 & 0.01297 & 53.6 & 17.85 & --15.90 & --15.27 & 8.01 & --1.45 & --9.46 & December 2016 & KOSMOS \\
151 & 167.20722 & 14.89958 & 0.01237 & 51.0 & 18.96 & --14.61 & --14.94 & 7.30 & --1.19 & --8.49 & April 2015 & KOSMOS \\
157 & 231.74008 & 11.19834 & 0.01301 & 55.8 & 18.95 & --14.87 & --14.09 & 7.43 & --1.89 & --9.32 & April 2015 & KOSMOS \\
%160 & 349.67995 & 26.22570 & 0.02077 & 92.3 & 18.54 & --16.47 & --15.40 & --- &  &  & December 2016 & KOSMOS \\
173 & 8.72176 & 24.60206 & 0.01932 & 86.1 & 18.67 & --16.08 & --15.08 & 7.94 & --1.22 & --9.16 & September 2014 & KOSMOS \\
174 & 26.69443 & 25.26602 & 0.01727 & 76.9 & 19.51 & --15.21 & --14.04 & 7.57 & --1.35 & --8.92 & December 2016 & KOSMOS \\
194 & 184.16381 & 14.29387 & 0.02386 & 103.0 & 17.83 & --17.32 & --16.34 & 8.67 & --0.91 & --9.58 & April 2015 & KOSMOS \\
218 & 348.58932 & 28.30260 & 0.02298 & 104.3 & 18.81 & --16.54 & --15.60 & 7.90 & --1.02 & --8.92 & October 2012 & RC Spec. \\
--- & --- & --- & --- & --- & --- & --- & --- & --- & --- & --- & September 2014 & KOSMOS \\
303 & 145.88482 & 33.44935 & 0.00188 & 12.1 & 19.43 & --11.02 & --10.75 & 5.55 & --2.93 & --8.48 & April 2015 & KOSMOS
\enddata
\tablecomments{\HA\ Dots 2, 81, and 218 were observed with both the RC Spectrograph and KOSMOS, and \HA\ Dot 43 was observed with KOSMOS during two separate runs.
Each source is therefore listed twice.}
%\tablecomments{$B$-band absolute magnitudes ($M_{B}$) calculated based on transformation relation of SDSS photometry from \citet{bib:Cook2014}.}
\end{deluxetable*}

\section{The Data} % Section 2.
\label{sec:data}

\begin{figure*} % Figure 1
\figurenum{1}
\plotone{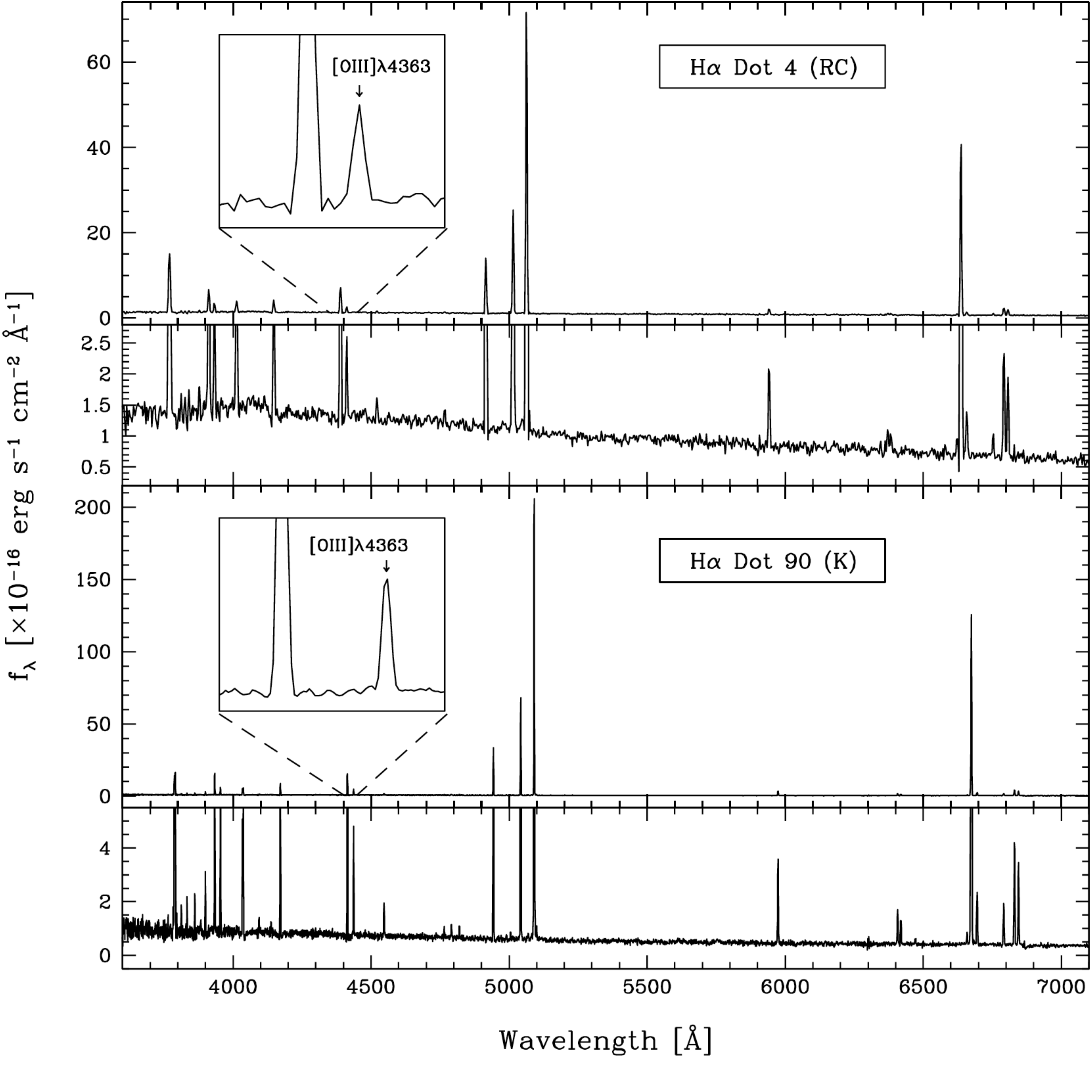}
\caption{
Optical spectra of two typical \HA\ Dots with \Te-method abundances taken from our sample.
Spectral emission-line features are representative of metal-poor star-forming systems.
The inset boxes highlight the \HG\ and \OIII$\lambda$4363 emission lines.
Vertical scaling has been reduced on the lower portions of the spectra to show increased detail for the weaker lines.
Instances in the spectra of particularly-strong absorption are likely instrumental artifacts (see \S\ref{sec:datareduction} for details).
\emph{Top}:\ \HA\ Dot 4 taken with the RC Spectrograph in April 2012.
\emph{Bottom}:\ \HA\ Dot 90 taken with KOSMOS in September 2014.
}
\label{fig:spectra_detail}
\end{figure*}

\begin{figure*} % Figure 2a
\figurenum{2}
\plotone{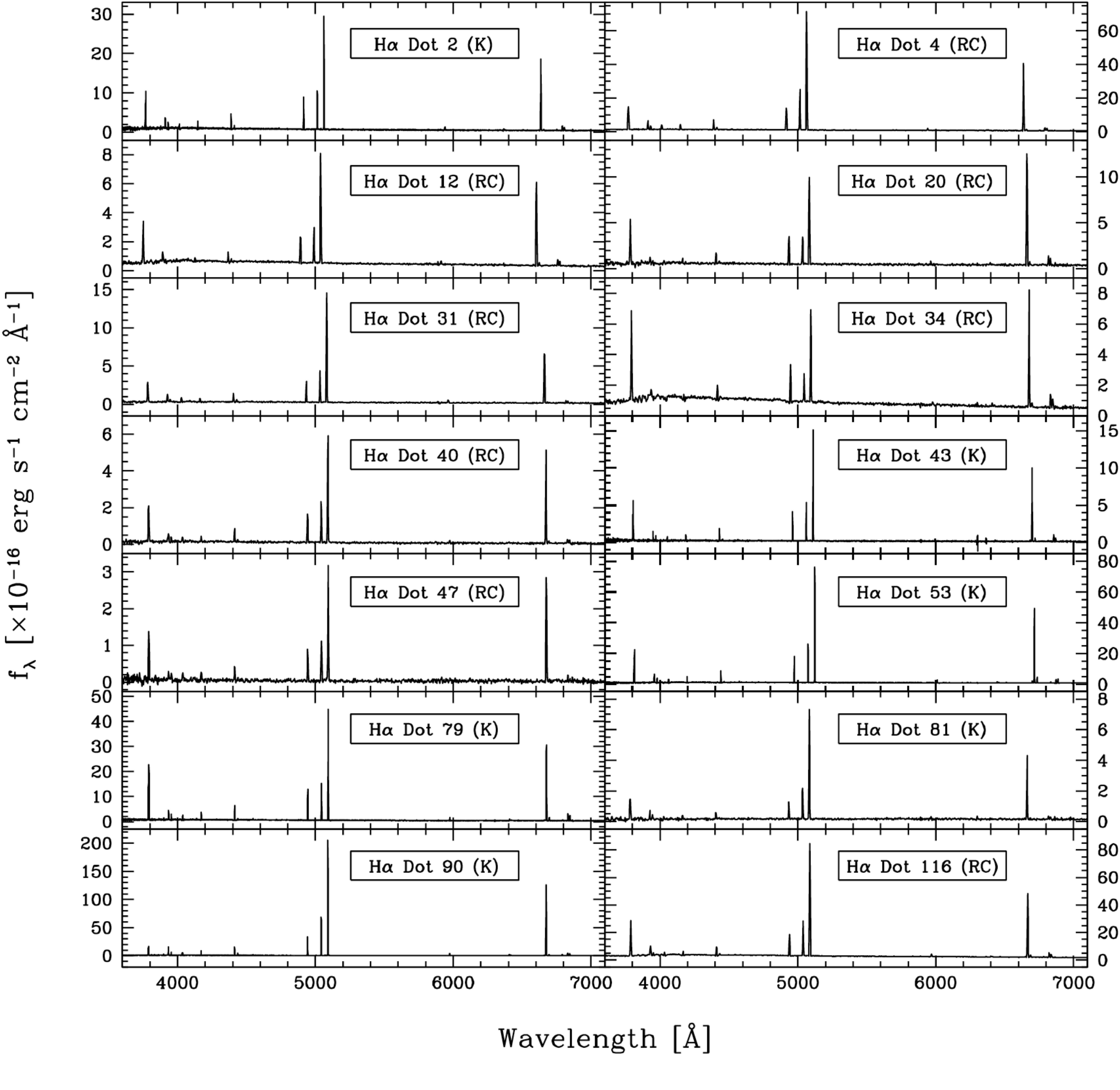}
\caption{
Optical spectra of \HA\ Dots from this study (1 of 2).
Sources observed with the RC Spectrograph are marked with ``(RC)", while those observed with KOSMOS are marked with ``(K)".
Three galaxies (\HA\ Dots 2, 81, and 218) were observed with both instruments, while one galaxy (\HA\ Dot 43) was observed with KOSMOS twice; only a single KOSMOS spectra for each are presented here.
}
\label{fig:spectra_all_1}
\end{figure*}

\begin{figure*} % Figure 2b
\figurenum{2}
\plotone{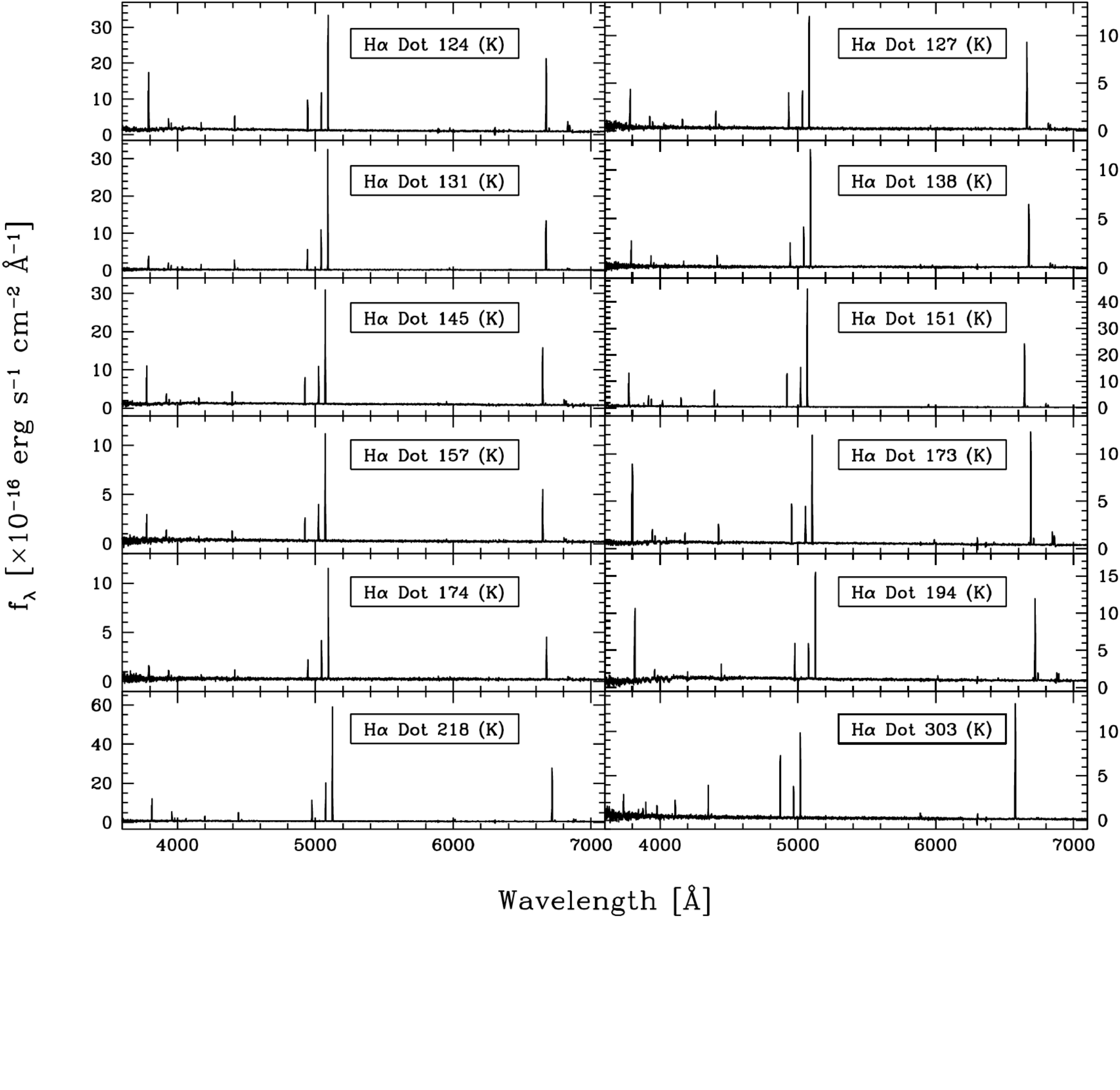}
\vspace{-55pt}
\caption{
Optical spectra of \HA\ Dots from this study (2 of 2).
Sources observed with the RC Spectrograph are marked with ``(RC)", while those observed with KOSMOS are marked with ``(K)".
Three galaxies (\HA\ Dots 2, 81, and 218) were observed with both instruments, while one galaxy (\HA\ Dot 43) was observed with KOSMOS twice; only a single KOSMOS spectra for each are presented here.
}
\label{fig:spectra_all_2}
\end{figure*}

\indent Spectroscopic observations of the \HA\ Dots presented in this study were undertaken at the Kitt Peak National Observatory (KPNO) Mayall 4m telescope over the course of five observing runs.
The first and second runs utilized the Richey--Chr\'{e}tien (RC) Focus Spectrograph, taking place in April and October 2012, respectively.
Metal abundances for another sample of dwarf galaxies observed during these two runs have been presented previously \citep{bib:Haurberg2015}.
The final three runs utilized the newer KPNO Ohio State Multi-Object Spectrograph (KOSMOS), taking place in September 2014, April 2015, and December 2016.
The number of \HA\ Dot spectra acquired from each run and presented in this study are, respectively; six in April 2012, two in October 2012, ten in September 2014, five in April 2015, and three in December 2016.
\\
\indent Three sources (\HA\ Dots 2, 81, and 218) were observed first with the RC Spectrograph, then later re-observed with KOSMOS, providing a direct comparison of the data between the two instruments.
One object (\HA\ Dot 43) was observed twice during separate KOSMOS runs.
In the case of these duplicated observations, the individual spectra were evaluated, and abundance analyses were carried out using only the better-quality of the two.
For \HA\ Dots 2, 81, and 218, this was the KOSMOS spectra, and for \HA\ Dot 43, this was the September 2014 data.
\\
\indent In total, we have obtained robust metallicities for twenty-six \HA\ Dots.
A summary of the \HA\ Dots covered in this study, including identifier, coordinates, redshift $z$, distance estimate, magnitude data (photometry), stellar mass estimate (as log $M_{\odot}$), star-formation rate (SFR; as $M_{\odot}\ yr^{-1}$), specific SFR (SFR per stellar mass; as log SFR $M_{\odot}^{-1}$), observing run, and spectroscopic instrument information, is presented in Table \ref{tab:Dot_info}.
Astrometry for these objects are adopted from SDSS \citep{bib:York2000, bib:Abazajian2004, bib:Alam2015}, while distance estimates are developed from Hubble Flow calculations using redshift information measured from these spectra\footnote{\citet{bib:Hirschauer2016} found that a distance estimation by the Hubble Flow method for \HA\ Dot 303 (also known as \Alecxy) was highly uncertain due to the presence of a local velocity anomaly along its line of sight.
We have instead adopted a distance of 12.1 Mpc as determined by \citet{bib:McQuinn2020} by the Tip of the Red Giant Branch (TRGB) method utilizing \emph{HST} photometry.}.
$R$-band magnitudes are derived from AHA photometry, and $B$-band magnitudes are calculated based on a transformation relation specialized for galaxies from \citet{bib:Cook2014} utilizing SDSS photometry,
\[
B - i = (1.27 \pm 0.03)(g - i) + (0.16 \pm 0.01),
\]
with $g$ and $i$ adopted from the SDSS.
The utility of these data for our sources are confirmed via consistency check between the AHA $R$-band apparent magnitudes and SDSS $r$-band magnitudes, which typically differ by no more than a few tenths of a magnitude.
% ALFALFA \HA\
We note that this relation's reliance on the $i$-band is not ideal, as the color term $(g - i)$ can be quite large and thus has high sensitivity to errors.
As a check of the derived $B$-band, we compare with the transformation of R.\ Lupton (2005)\footnote{\url{https://www.sdss3.org/dr10/algorithms/sdssUBVRITransform.php}}, which is based on stellar photometry,
\[
B = g + 0.3130(g - r) + 0.2271,
\]
finding near-total agreement.
The derivation of the remaining quantities in Table \ref{tab:Dot_info}, stellar mass, star-formation rate (SFR), and specific star-formation rate (sSFR), are described in Section \ref{sec:MassAndSFR}.
%Finally, stellar mass estimates were obtained using SED fitting techniques utilizing photometry from...
%\texttt{Steven: What photometry was utilized in the SED fitting?
%And what was the code, probably CIGALE?}

\subsection{Richey--Chr\'{e}tien Focus Spectrographic Data} % Section 2.1.
\label{sec:RCSpec}
% 18 June 2020
% NB I am taking this text from Haurberg+ (2015) since I assume the observing setup is roughly the same

\indent Spectral observations taken during the first two runs were carried out using the RC Spectrograph and T2KA detector on the Mayall 4m at KPNO.
The KPC-10A grating (316 lines mm$^{-1}$) and WG-345 blocking filter were used.
The grating is blazed at 4000 \AA\ giving a dispersion of 2.78 \AA\ pixel$^{-1}$ and total coverage from 2850-8550 \AA\ on the CCD, allowing for recovery of all diagnostic emission lines required for abundance work in a single instrumental setup.
All spectra were taken with a slit width of 1$\arcsec$.5 and the slit extended 342$\arcsec$ along the spatial direction.
The instrumental pixel scale is 0.69 arcsec pixel$^{-1}$ in the spatial direction\footnote{\url{https://www.noao.edu/kpno/manuals/rcspec/rcsp.html}}. 
\\
\indent This instrumental setup delivered an effective spectral resolution of $\sim$5.6 \AA.
Most of the data were obtained under clear sky conditions, and with seeing ranging from 0$\arcsec$.8 to 1$\arcsec$.5.
Each source was observed for three different exposures of 1200 seconds which were later combined into one final image with a total exposure time of 3600 seconds.
The slit was positioned along the parallactic angle as closely as possible in an effort to avoid the effects of differential atmospheric diffraction \citep{bib:Filippenko1982}.

\subsection{KOSMOS Spectrographic Data} % Section 2.2.
\label{sec:KOSMOS}

\indent Spectral observations taken during the final three runs were carried out using KOSMOS, which replaced the RC Spectrograph on the KPNO Mayall 4m in 2014.
Each \HA\ Dot was observed first with a blue-sensitive grating and blue-sensitive slit instrumental setup, which provided wavelength coverage of $\sim$3500-6200 \AA.
% , sufficient for detecting the important \OII, \OIII, and hydrogen Balmer recombination lines
Next, the grating was switched to one that is red-sensitive while keeping the same slit (avoiding the need to re-position the telescope), which covered a wavelength range of $\sim$5000-9000 \AA.
Both the blue- and red-sensitive gratings possess a spectral resolution of R $\sim$ 2100\footnote{\url{https://www.noao.edu/kpno/manuals/kosmosman/KOSMOS-Manual.pdf}}.
For these low-redshift systems, our observing strategy recovered all optical-band emission lines necessary for abundance work with no gap in spectral coverage.
\\
\indent The instrumental pixel scale is 0.29 arcsec pixel$^{-1}$ in the spatial direction.
Dispersion for the blue spectral setup is 0.66 \AA\ pixel$^{-1}$, while for the red it is 0.99 \AA\ pixel$^{-1}$.
Spectra were taken with a slit width of 1$\arcsec$.2; the slit extents 10$\arcmin$ in the spatial direction.
Sky conditions were generally clear at the time of the observations.
Image quality was derived from the measurement of stellar profiles in the object acquisition images, with FWHM values ranging between 0$\arcsec$.9 and 2$\arcsec$.0, but typically equalling $\sim$1$\arcsec$.2, equivalent to the slit width.
Observing strategy generally consisted of three exposures using the blue setup and two exposures using the red setup, each of which were 900 seconds in length.
As with observations taken with the RC Spectrograph, the effects of differential atmospheric diffraction were mitigated by aligning the position angle of the slit to the parallactic angle.

\subsection{Data Reduction and Emission Line Measurement} % Section 2.3.
\label{sec:datareduction}

\indent The data reduction for both the KPNO 4m RC Spectrograph and KOSMOS observations were carried out with the Image Reduction and Analysis Facility (\texttt{IRAF}).
%(\texttt{IRAF})\footnote{\texttt{IRAF} was the Image Reduction and Analysis Facility distributed by the National Optical Astronomy Observatory (NOAO), which was operated by the Association of Universities for Research in Astronomy (AURA) under cooperative agreement with the National Science Foundation (NSF).}.
For both data sets all of the reduction steps mentioned below were carried out on the individual spectral images independently.
The fully reduced 1D spectra were then combined into a single high signal-to-noise ratio (S/N) spectrum prior to the measurement stage.
\\
\indent Processing of the two-dimensional spectral images followed standard methods.
The bias level was determined and subtracted from each image using the overscan region.
A mean bias image was then created by combining 10 zero-second exposures taken on each night of observation.
This image was subtracted to correct the science images for any possible two-dimensional structure in the bias.
Flat-fielding was achieved using an average-combined quartz lamp image that was corrected for the wavelength-dependent response of the system.
For cosmic ray rejection, we used \texttt{L.A.Cosmic} \citep{bib:vanDokkum2001}, taking special care that no emission lines were ``clipped" by the software by visually inspecting the image results.
\\
\indent One-dimensional spectra were extracted using the \texttt{IRAF APALL} routine.
The extraction widths (i.e., distances along the slit) were set on an individual basis depending on the presentation of the source, typically ranging between 3 and 7 arcsec for both RC Spec and KOSMOS spectra, with a typical value of 5.5 arcsec.
The extractions were conservatively wide such that modest changes in the extraction widths resulted in negligible changes in the spectra.
Narrower spectral extractions can achieve higher signal-to-noise in the faintest lines, but at a loss of spectral fidelity, in the sense that some flux is being missed.
Sky subtraction was also performed at this stage, with the sky spectrum being measured in regions on either side of the object extraction window.
HeNeAr lamp spectra were used to assign a wavelength scale, and the spectra of spectrophotometric standard stars, including BD+17 4708, HD 84937, HD 19445 \citep{bib:OkeGunn1983}, Feige 34 \citep{bib:Massey1988}, BD+28 4211, G191-B2B, and Hz 44 \citep{bib:Oke1990} were used to establish the flux scale.
\\
\indent Fully reduced \HA\ Dot spectra are shown in Figures \ref{fig:spectra_detail} and \ref{fig:spectra_all_1}.
Figure \ref{fig:spectra_detail} presents \HA\ Dot 4, taken with the RC Spectrograph (top), and \HA\ Dot 90, taken with KOSMOS (bottom).
Inset boxes highlight the \HG\ and \OIII$\lambda$4363 emission lines, the latter being critical for determining the temperature of the electron gas ($T_{e}$).
Figure \ref{fig:spectra_all_1} displays spectra for all \HA\ Dots from this study.
Those observed with the RC Spectrograph are labeled with ``(RC)", while those observed with KOSMOS are labeled with ``(K)".
\\
\indent KOSMOS spectra are obtained with both a blue- and red-sensitive setup, with both independent spectra plotted together in Figure \ref{fig:spectra_detail}.
The blue and red spectra overlap between $\sim$5000 and 6200 \AA, which is sufficient such that there is no break in spectral coverage.
Extraction regions for the blue and red spectra were carefully matched to ensure that the same portion of the galaxy was being extracted.
We checked to ensure that the flux scales in both spectral regions agreed by comparing the fluxes of the \OIII$\lambda$5007 line that was located in both the blue and red spectra for our instrumental setup.
The flux of this line on the blue side was characteristically 5\% higher than that of the red side.
We therefore scaled the flux of the red spectra up by the measured ratio of the blue to red line fluxes in order to account for this difference.
In cases for which the \OIII$\lambda$5007 line was not measured, or the relative line flux was severely impacted by the sensitivity function at the extreme blueward end of the red-side spectra, an average scaling of 5\% was applied.
The only effect that this flux scaling has on our results for the oxygen abundances of the \HA\ Dots is on the determination of the reddening parameter \cHB\ based on the \HA/\HB\ line ratio.
Finally, we corrected the spectra for telluric absorption by using the spectra of our spectrophotometric standard stars.
The primary impact of this correction was to improve the measurement of the \SII$\lambda\lambda$6717,6731 lines for galaxies whose redshifts moved these lines into the $B$-band.
\\
%
%\indent For both the RC Spectrograph and KOSMOS spectra, the emission lines were measured using the \texttt{IRAF} routine \texttt{SPLOT} by directly summing the flux under each line.
%\texttt{UPDATE: Spectral lines were done with WRALF!}
\indent Measurement of the line fluxes was carried out using \texttt{WRALF} (WRapped Automated Line Fitting; \citealp{bib:Cousins2019}), which was developed to automate the process of measuring the spectral features in emission-line galaxies.
\texttt{WRALF} is a Python code that serves as a wrapper for \texttt{ALFA} (Automated Line Fitting Algorithm; \citealp{bib:Wesson2016}).
\texttt{ALFA} identifies all emission lines in a spectrum, determines the local continuum, then fits each line with a Gaussian profile.
\texttt{WRALF} returns line positions, integrated line fluxes and uncertainties, equivalent widths, and line widths for each emission feature found in the spectrum.
Before applying \texttt{WRALF} to the \HA\ Dot spectra, we carried out tests to verify the quality of the flux measurements.
Measurements obtained with \texttt{WRALF} agreed extremely well with fluxes measured by hand using \texttt{IRAF} \texttt{SPLOT}.
\\
\indent Internal reddening for each source was found by calculating \cHB\ determined from simultaneous fits to the reddening and underlying absorption in the Balmer lines.
Under the simplifying assumption that the same equivalent width (EW) of underlying absorption applies to each Balmer line, an absorption line correction was applied to the spectrum that ranged in value from 0 to 5 \AA.
The value for the underlying absorption was varied in 0.5 \AA\ increments until a self-consistent value of \cHB\ for the Balmer line ratios \HA/\HB, \HG/\HB, and \HD/\HB\ was found.
This process led to the determination of characteristic values for the underlying absorption, which typically ranged from 0.5 to 2.0 \AA.
Because our target sources are dwarf galaxies, which generally have very low internal velocity dispersion from their stellar component, the width of any underlying absorption is typically narrower than the instrumental resolution.
This absorption is therefore filled in by the emission lines, hiding it from direct observation.
We adopt the standard practice of utilizing multiple Balmer lines to infer the underlying absorption simultaneously with an appropriate value of the reddening coefficient, allowing for the correction of this absorption even in cases when it is not directly observed.
\\
\indent The value of \cHB\ is then used to correct the measured line ratios for reddening, following the standard procedure (e.g., \citealp{bib:OsterbrockFerland2006}),
\[
\frac{I(\lambda)}{I(\text{H}\beta)} = \frac{I_{\lambda 0}}{I_{\text{H}\beta 0}}\ 10^{-c_{\text{H}\beta} [f(\lambda) - f(\text{H}\beta)]},
\]
\noindent where $f(\lambda$) is derived from studies of absorption in the Milky Way (using values taken from \citealp{bib:Rayo1982}, which are derived from the extinction law of \citealp{bib:Whitford1958}).
In the cases where flux scaling was applied between the KOSMOS blue- and red-sensitive setups, any effect of the \HA/\HB\ ratio on \cHB\ is mitigated by utilizing the \HG/\HB\ and \HD/\HB\ ratios, which are derived from the blue spectra only.
For the three \HA\ Dots that were observed with both RC Spec and KOSMOS (\HA\ Dots 2, 81, and 218), the values of \cHB\ derived with the RC Spec data show excellent agreement with the KOSMOS values.
Similarly, the \cHB\ values for \HA\ Dot 43, which was observed on two separate occasions with KOSMOS, are effectively identical.

\begin{deluxetable*}{ccccccccccc} % Table 2a
\tablenum{2}
\label{tab:ratios_alla}
\rotate
\tabletypesize{\scriptsize}
\tablewidth{0pt}
\tablecaption{
Emission-line ratios relative to \HB\ (part 1 of 3).
}
\tablehead{
\colhead{Ion}&\colhead{$\lambda$ [\AA]}&\colhead{\HA\ Dot 2 (K)}&\colhead{\HA\ Dot 4 (RC)}&\colhead{\HA\ Dot 12 (RC)}&\colhead{\HA\ Dot 20 (RC)}&\colhead{\HA\ Dot 31 (RC)}&\colhead{\HA Dot 34 (RC)}&\colhead{\HA\ Dot 40 (RC)}&\colhead{\HA\ Dot 43 (K)}&\colhead{\HA\ Dot 47 (RC)}
}
\startdata
\OII\	&	3727.43	&	1.7509 $\pm$ 0.0619	&	1.2188 $\pm$ 0.0399	&	1.3672 $\pm$ 0.0723	&	2.2637 $\pm$ 0.1096	&	1.0863 $\pm$ 0.0320	&	2.4336 $\pm$ 0.1067	&	1.4519 $\pm$ 0.0603	&	2.4673 $\pm$ 0.0813	&	1.9161 $\pm$ 0.1388	\\
H 10	&	3797.90	&	$\ldots$	&	$\ldots$	&	$\ldots$	&	$\ldots$	&	$\ldots$	&	$\ldots$	&	$\ldots$	&	$\ldots$	&	$\ldots$	\\
H 9	&	3835.39	&	$\ldots$	&	$\ldots$	&	$\ldots$	&	$\ldots$	&	$\ldots$	&	$\ldots$	&	$\ldots$	&	$\ldots$	&	$\ldots$	\\
\NeIII\	&	3868.74	&	0.2993 $\pm$ 0.0132	&	0.3831 $\pm$ 0.0132	&	0.2690 $\pm$ 0.0283	&	0.1992 $\pm$ 0.0258	&	0.3365 $\pm$ 0.0176	&	0.1717 $\pm$ 0.0464	&	0.2487 $\pm$ 0.0229	&	0.3379 $\pm$ 0.0172	&	0.2924 $\pm$ 0.0398	\\
He  I + H 8	&	3888.65	&	0.1340 $\pm$ 0.0103	&	0.1380 $\pm$ 0.0088	&	$\ldots$	&	$\ldots$	&	0.1066 $\pm$ 0.0156	&	$\ldots$	&	0.1487 $\pm$ 0.0234	&	0.1763 $\pm$ 0.0140	&	0.1933 $\pm$ 0.0395	\\
\NeIII\	&	3967.47	&	0.0427 $\pm$ 0.0105	&	0.1530 $\pm$ 0.0068	&	$\ldots$	&	$\ldots$	&	0.0880 $\pm$ 0.0162	&	$\ldots$	&	$\ldots$	&	0.0675 $\pm$ 0.0102	&	$\ldots$	\\
\HE\	&	3970.08	&	0.0851 $\pm$ 0.0106	&	0.0648 $\pm$ 0.0062	&	$\ldots$	&	$\ldots$	&	0.1085 $\pm$ 0.0162	&	$\ldots$	&	0.1103 $\pm$ 0.0241	&	0.1337 $\pm$ 0.0103	&	0.1877 $\pm$ 0.0382	\\
He I	&	4026.19	&	$\ldots$	&	$\ldots$	&	$\ldots$	&	$\ldots$	&	0.0600 $\pm$ 0.0162	&	$\ldots$	&	$\ldots$	&	$\ldots$	&	$\ldots$	\\
\HD\	&	4101.74	&	0.2566 $\pm$ 0.0142	&	0.2384 $\pm$ 0.0080	&	0.2111 $\pm$ 0.0146	&	0.2664 $\pm$ 0.0277	&	0.1646 $\pm$ 0.0114	&	0.2431 $\pm$ 0.0131	&	0.1910 $\pm$ 0.0182	&	0.2627 $\pm$ 0.0133	&	0.2798 $\pm$ 0.0428	\\
\HG\	&	4340.47	&	0.4731 $\pm$ 0.0121	&	0.4839 $\pm$ 0.0098	&	0.4329 $\pm$ 0.0236	&	0.4809 $\pm$ 0.0187	&	0.4200 $\pm$ 0.0163	&	0.4745 $\pm$ 0.0275	&	0.4771 $\pm$ 0.0180	&	0.4726 $\pm$ 0.0118	&	0.4773 $\pm$ 0.0336	\\
\OIII\	&	4363.21	&	0.0816 $\pm$ 0.0059	&	0.0878 $\pm$ 0.0068	&	0.0766 $\pm$ 0.0410	&	0.0792 $\pm$ 0.0160	&	0.0932 $\pm$ 0.0137	&	0.0531 $\pm$ 0.0076	&	0.0702 $\pm$ 0.0151	&	0.0520 $\pm$ 0.0049	&	0.0992 $\pm$ 0.0242	\\
He I	&	4471.48	&	0.0262 $\pm$ 0.0030	&	0.0340 $\pm$ 0.0051	&	$\ldots$	&	$\ldots$	&	$\ldots$	&	$\ldots$	&	$\ldots$	&	0.0438 $\pm$ 0.0057	&	$\ldots$	\\
He II	&	4685.71	&	0.0255 $\pm$ 0.0024	&	$\ldots$	&	$\ldots$	&	$\ldots$	&	$\ldots$	&	$\ldots$	&	$\ldots$	&	$\ldots$	&	$\ldots$	\\
\ArIV\	&	4711.26	&	$\ldots$	&	$\ldots$	&	$\ldots$	&	$\ldots$	&	$\ldots$	&	$\ldots$	&	$\ldots$	&	$\ldots$	&	$\ldots$	\\
\ArIV\	&	4740.12	&	$\ldots$	&	$\ldots$	&	$\ldots$	&	$\ldots$	&	$\ldots$	&	$\ldots$	&	$\ldots$	&	$\ldots$	&	$\ldots$	\\
\HB\	&	4861.33	&	1.0000 $\pm$ 0.0162	&	1.0000 $\pm$ 0.0069	&	1.0000 $\pm$ 0.0164	&	1.0000 $\pm$ 0.0206	&	1.0000 $\pm$ 0.0161	&	1.0000 $\pm$ 0.0244	&	1.0000 $\pm$ 0.0216	&	1.0000 $\pm$ 0.0120	&	1.0000 $\pm$ 0.0425	\\
He I	&	4921.93	&	$\ldots$	&	$\ldots$	&	$\ldots$	&	$\ldots$	&	$\ldots$	&	$\ldots$	&	$\ldots$	&	$\ldots$	&	$\ldots$	\\
\OIII\	&	4958.91	&	1.1100 $\pm$ 0.0140	&	1.7393 $\pm$ 0.0122	&	1.1467 $\pm$ 0.0212	&	0.9068 $\pm$ 0.0213	&	1.7410 $\pm$ 0.0263	&	0.7195 $\pm$ 0.0204	&	1.4544 $\pm$ 0.0331	&	1.2692 $\pm$ 0.0120	&	1.1985 $\pm$ 0.0550	\\
\OIII\	&	5006.84	&	3.2832 $\pm$ 0.0394	&	5.1710 $\pm$ 0.0304	&	3.5008 $\pm$ 0.0442	&	2.7400 $\pm$ 0.0439	&	5.3010 $\pm$ 0.0636	&	2.2059 $\pm$ 0.0412	&	3.7834 $\pm$ 0.0620	&	3.8354 $\pm$ 0.0359	&	3.7622 $\pm$ 0.1206	\\
\NI\	&	5197.90	&	$\ldots$	&	$\ldots$	&	$\ldots$	&	$\ldots$	&	$\ldots$	&	$\ldots$	&	$\ldots$	&	$\ldots$	&	$\ldots$	\\
He I	&	5875.62	&	0.0952 $\pm$ 0.0042	&	0.0889 $\pm$ 0.0044	&	0.0949 $\pm$ 0.0105	&	0.0871 $\pm$ 0.0135	&	0.1332 $\pm$ 0.0129	&	0.0662 $\pm$ 0.0147	&	0.0742 $\pm$ 0.0110	&	0.1016 $\pm$ 0.0065	&	$\ldots$	\\
\OI\	&	6300.30	&	0.0470 $\pm$ 0.0066	&	0.0215 $\pm$ 0.0029	&	$\ldots$	&	0.0569 $\pm$ 0.0082	&	$\ldots$	&	0.0724 $\pm$ 0.0100	&	0.0465 $\pm$ 0.0114	&	0.0446 $\pm$ 0.0041	&	$\ldots$	\\
\SIII\	&	6312.06	&	$\ldots$	&	0.0178 $\pm$ 0.0029	&	$\ldots$	&	$\ldots$	&	$\ldots$	&	$\ldots$	&	$\ldots$	&	0.0184 $\pm$ 0.0044	&	$\ldots$	\\
\OI\	&	6363.78	&	$\ldots$	&	$\ldots$	&	$\ldots$	&	$\ldots$	&	$\ldots$	&	$\ldots$	&	$\ldots$	&	$\ldots$	&	$\ldots$	\\
\NII\	&	6548.05	&	$\ldots$	&	0.0148 $\pm$ 0.0042	&	$\ldots$	&	$\ldots$	&	$\ldots$	&	$\ldots$	&	$\ldots$	&	0.0424 $\pm$ 0.0062	&	$\ldots$	\\
\HA\	&	6562.82	&	2.7744 $\pm$ 0.0958	&	2.7999 $\pm$ 0.0867	&	2.7823 $\pm$ 0.0972	&	2.7626 $\pm$ 0.1054	&	2.7571 $\pm$ 0.0333	&	2.7759 $\pm$ 0.1085	&	2.7959 $\pm$ 0.1049	&	2.8147 $\pm$ 0.0928	&	2.7690 $\pm$ 0.1484	\\
\NII\	&	6583.46	&	0.0471 $\pm$ 0.0041	&	0.0448 $\pm$ 0.0037	&	0.0827 $\pm$ 0.0139	&	0.0796 $\pm$ 0.0140	&	0.0445 $\pm$ 0.0109	&	0.1072 $\pm$ 0.0119	&	0.0762 $\pm$ 0.0129	&	0.1235 $\pm$ 0.0058	&	0.0596 $\pm$ 0.0193	\\
He I	&	6678.15	&	0.0230 $\pm$ 0.0027	&	0.0239 $\pm$ 0.0022	&	$\ldots$	&	$\ldots$	&	$\ldots$	&	$\ldots$	&	$\ldots$	&	0.0301 $\pm$ 0.0046	&	$\ldots$	\\
\SII\	&	6716.44	&	0.1640 $\pm$ 0.0072	&	0.1185 $\pm$ 0.0049	&	0.1946 $\pm$ 0.0105	&	0.2279 $\pm$ 0.0139	&	0.1159 $\pm$ 0.0104	&	0.3311 $\pm$ 0.0208	&	0.1260 $\pm$ 0.0133	&	0.2418 $\pm$ 0.0109	&	0.1537 $\pm$ 0.0213	\\
\SII\	&	6730.81	&	0.1218 $\pm$ 0.0055	&	0.0676 $\pm$ 0.0039	&	0.1402 $\pm$ 0.0091	&	0.1469 $\pm$ 0.0126	&	0.1007 $\pm$ 0.0104	&	0.2124 $\pm$ 0.0199	&	0.1053 $\pm$ 0.0128	&	0.1741 $\pm$ 0.0090	&	$\ldots$	\\
 He I	&	7065.20	&	0.0215 $\pm$ 0.0026	&	0.0224 $\pm$ 0.0029	&	$\ldots$	&	$\ldots$	&	$\ldots$	&	$\ldots$	&	$\ldots$	&	0.0229 $\pm$ 0.0045	&	$\ldots$	\\
\ArIII\	&	7135.79	&	0.0474 $\pm$ 0.0043	&	0.0607 $\pm$ 0.0037	&	0.0606 $\pm$ 0.0128	&	0.0554 $\pm$ 0.0135	&	0.0679 $\pm$ 0.0094	&	$\ldots$	&	$\ldots$	&	0.0903 $\pm$ 0.0084	&	$\ldots$	\\
\OII\	&	7318.99	&	$\ldots$	&	0.0210 $\pm$ 0.0038	&	$\ldots$	&	$\ldots$	&	$\ldots$	&	0.0534 $\pm$ 0.0155	&	$\ldots$	&	0.0289 $\pm$ 0.0056	&	$\ldots$	\\
\OII\	&	7329.73	&	0.0126 $\pm$ 0.0030	&	$\ldots$	&	$\ldots$	&	0.0497 $\pm$ 0.0120	&	$\ldots$	&	$\ldots$	&	$\ldots$	&	0.0248 $\pm$ 0.0060	&	$\ldots$	\\
\hline																					
c$_{\text{H}\beta}$	&		&	0.155 $\pm$ 0.042	&	0.133 $\pm$ 0.040	&	0.024 $\pm$ 0.042	&	0.420 $\pm$ 0.044	&	0.00 $\pm$ ---	&	0.030 $\pm$ 0.045	&	0.085 $\pm$ 0.044	&	0.253 $\pm$ 0.041	&	0.288 $\pm$ 0.056	\\
EW(\HB)	&		&	31.64 \AA\	&	73.92 \AA\	&	27.72 \AA\	&	39.70 \AA\	&	57.30 \AA\	&	19.68 \AA\	&	79.60 \AA\	&	46.69 \AA\	&	140.33 \AA\	\\
F(\HB)*	&		&	2.4439 $\pm$ 0.0279	&	8.6630 $\pm$ 0.0423	&	1.3720 $\pm$ 0.0159	&	2.1221 $\pm$ 0.0308	&	1.6180 $\pm$ 0.0184	&	1.6281 $\pm$ 0.0281	&	0.9917 $\pm$ 0.0151	&	1.0233 $\pm$ 0.0087	&	0.5459 $\pm$ 0.0164	
\enddata
\tablecomments{Line flux of \HB\ in units of 10$^{-15}$ erg s$^{-1}$ cm$^{-2}$.}
\end{deluxetable*}

\begin{deluxetable*}{ccccccccccc} % Table 2b
\tablenum{2}
\label{tab:ratios_allb}
\rotate
\tabletypesize{\scriptsize}
\tablewidth{0pt}
\tablecaption{
Emission-line ratios relative to \HB\ (part 2 of 3).
}
\tablehead{
\colhead{Ion}&\colhead{$\lambda$ [\AA]}&\colhead{\HA\ Dot 53 (K)}&\colhead{\HA\ Dot 79 (K)}&\colhead{\HA\ Dot 81 (K)}&\colhead{\HA\ Dot 90 (K)}&\colhead{\HA\ Dot 116 (RC)}&\colhead{\HA\ Dot 124 (K)}&\colhead{\HA\ Dot 127 (K)}&\colhead{\HA\ Dot 131 (K)}&\colhead{\HA\ Dot 138 (K)}
}
\startdata
\OII\	&	3727.43	&	2.6170 $\pm$ 0.0842	&	3.0867 $\pm$ 0.0949	&	1.5054 $\pm$ 0.0475	&	1.0749 $\pm$ 0.0372	&	1.9625 $\pm$ 0.0684	&	3.2592 $\pm$ 0.1272	&	2.1355 $\pm$ 0.0913	&	1.2027 $\pm$ 0.0530	&	1.9615 $\pm$ 0.0991	\\
H 10	&	3797.90	&	$\ldots$	&	$\ldots$	&	0.0425 $\pm$ 0.0075	&	0.0441 $\pm$ 0.0053	&	$\ldots$	&	$\ldots$	&	$\ldots$	&	$\ldots$	&	$\ldots$	\\
H 9	&	3835.39	&	$\ldots$	&	0.0463 $\pm$ 0.0082	&	0.0637 $\pm$ 0.0105	&	0.0657 $\pm$ 0.0064	&	$\ldots$	&	$\ldots$	&	$\ldots$	&	$\ldots$	&	$\ldots$	\\
\NeIII\	&	3868.74	&	0.3596 $\pm$ 0.0134	&	0.3107 $\pm$ 0.0098	&	0.5036 $\pm$ 0.0151	&	0.5268 $\pm$ 0.0139	&	0.4024 $\pm$ 0.0243	&	0.3206 $\pm$ 0.0158	&	0.3493 $\pm$ 0.0182	&	0.3558 $\pm$ 0.0160	&	0.4258 $\pm$ 0.0228	\\
He  I + H 8	&	3888.65	&	0.1657 $\pm$ 0.0094	&	0.1800 $\pm$ 0.0090	&	0.2171 $\pm$ 0.0089	&	0.1998 $\pm$ 0.0067	&	0.0843 $\pm$ 0.0205	&	0.1601 $\pm$ 0.0145	&	0.1889 $\pm$ 0.0200	&	0.2094 $\pm$ 0.0121	&	0.1916 $\pm$ 0.0232	\\
\NeIII\	&	3967.47	&	$\ldots$	&	0.0673 $\pm$ 0.0109	&	0.1256 $\pm$ 0.0067	&	0.1406 $\pm$ 0.0054	&	0.0751 $\pm$ 0.0164	&	$\ldots$	&	$\ldots$	&	0.1233 $\pm$ 0.0119	&	$\ldots$	\\
\HE\	&	3970.08	&	0.1161 $\pm$ 0.0121	&	0.1351 $\pm$ 0.0107	&	0.1589 $\pm$ 0.0072	&	0.1705 $\pm$ 0.0058	&	$\ldots$	&	$\ldots$	&	0.1318 $\pm$ 0.0232	&	0.1501 $\pm$ 0.0132	&	0.1230 $\pm$ 0.0266	\\
He I	&	4026.19	&	0.0218 $\pm$ 0.0045	&	0.0145 $\pm$ 0.0024	&	0.0278 $\pm$ 0.0046	&	0.0180 $\pm$ 0.0020	&	$\ldots$	&	$\ldots$	&	$\ldots$	&	$\ldots$	&	$\ldots$	\\
\HD\	&	4101.74	&	0.2638 $\pm$ 0.0135	&	0.2663 $\pm$ 0.0103	&	0.2658 $\pm$ 0.0091	&	0.2696 $\pm$ 0.0067	&	0.2466 $\pm$ 0.0099	&	0.2685 $\pm$ 0.0213	&	0.2768 $\pm$ 0.0129	&	0.2655 $\pm$ 0.0138	&	0.2679 $\pm$ 0.0162	\\
\HG\	&	4340.47	&	0.4735 $\pm$ 0.0112	&	0.4945 $\pm$ 0.0094	&	0.4992 $\pm$ 0.0084	&	0.4944 $\pm$ 0.0076	&	0.4799 $\pm$ 0.0158	&	0.4711 $\pm$ 0.0208	&	0.4885 $\pm$ 0.0180	&	0.4833 $\pm$ 0.0111	&	0.4801 $\pm$ 0.0211	\\
\OIII\	&	4363.21	&	0.0424 $\pm$ 0.0037	&	0.0454 $\pm$ 0.0044	&	0.0780 $\pm$ 0.0023	&	0.1361 $\pm$ 0.0026	&	0.0618 $\pm$ 0.0084	&	0.0464 $\pm$ 0.0079	&	0.0891 $\pm$ 0.0126	&	0.0827 $\pm$ 0.0064	&	0.0881 $\pm$ 0.0186	\\
He I	&	4471.48	&	0.0383 $\pm$ 0.0025	&	0.0374 $\pm$ 0.0033	&	0.0470 $\pm$ 0.0035	&	0.0411 $\pm$ 0.0017	&	0.0390 $\pm$ 0.0053	&	0.0395 $\pm$ 0.0089	&	0.0557 $\pm$ 0.0124	&	0.0452 $\pm$ 0.0060	&	$\ldots$	\\
He II	&	4685.71	&	$\ldots$	&	$\ldots$	&	$\ldots$	&	0.0115 $\pm$ 0.0012	&	0.0152 $\pm$ 0.0042	&	$\ldots$	&	$\ldots$	&	$\ldots$	&	$\ldots$	\\
\ArIV\	&	4711.26	&	$\ldots$	&	$\ldots$	&	$\ldots$	&	0.0156 $\pm$ 0.0013	&	$\ldots$	&	$\ldots$	&	$\ldots$	&	$\ldots$	&	$\ldots$	\\
\ArIV\	&	4740.12	&	$\ldots$	&	$\ldots$	&	0.0082 $\pm$ 0.0018	&	0.0117 $\pm$ 0.0018	&	0.0148 $\pm$ 0.0041	&	$\ldots$	&	$\ldots$	&	$\ldots$	&	$\ldots$	\\
\HB\	&	4861.33	&	1.0000 $\pm$ 0.0124	&	1.0000 $\pm$ 0.0097	&	1.0000 $\pm$ 0.0046	&	1.0000 $\pm$ 0.0031	&	1.0000 $\pm$ 0.0166	&	1.0000 $\pm$ 0.0219	&	1.0000 $\pm$ 0.0179	&	1.0000 $\pm$ 0.0136	&	1.0000 $\pm$ 0.0180	\\
He I	&	4921.93	&	$\ldots$	&	$\ldots$	&	0.0091 $\pm$ 0.0015	&	0.0098 $\pm$ 0.0019	&	$\ldots$	&	$\ldots$	&	$\ldots$	&	$\ldots$	&	$\ldots$	\\
\OIII\	&	4958.91	&	1.3964 $\pm$ 0.0129	&	1.1511 $\pm$ 0.0096	&	1.9628 $\pm$ 0.0081	&	2.0663 $\pm$ 0.0062	&	1.6879 $\pm$ 0.0226	&	1.1651 $\pm$ 0.0203	&	1.0552 $\pm$ 0.0201	&	1.9488 $\pm$ 0.0206	&	1.5703 $\pm$ 0.0235	\\
\OIII\	&	5006.84	&	4.2480 $\pm$ 0.0396	&	3.5206 $\pm$ 0.0266	&	5.9943 $\pm$ 0.0261	&	6.2247 $\pm$ 0.0228	&	5.0505 $\pm$ 0.0619	&	3.5446 $\pm$ 0.0570	&	3.1465 $\pm$ 0.0442	&	5.8887 $\pm$ 0.0601	&	4.6829 $\pm$ 0.0647	\\
\NI\	&	5197.90	&	0.0107 $\pm$ 0.0020	&	0.0105 $\pm$ 0.0021	&	$\ldots$	&	0.0030 $\pm$ 0.0008	&	$\ldots$	&	$\ldots$	&	$\ldots$	&	0.0164 $\pm$ 0.0043	&	$\ldots$	\\
He I	&	5875.62	&	0.1091 $\pm$ 0.0044	&	0.1016 $\pm$ 0.0036	&	0.1024 $\pm$ 0.0039	&	0.1095 $\pm$ 0.0027	&	0.1085 $\pm$ 0.0049	&	0.1027 $\pm$ 0.0080	&	0.0807 $\pm$ 0.0073	&	0.0877 $\pm$ 0.0054	&	0.0937 $\pm$ 0.0119	\\
\OI\	&	6300.30	&	0.0522 $\pm$ 0.0023	&	0.0468 $\pm$ 0.0025	&	0.0225 $\pm$ 0.0016	&	0.0255 $\pm$ 0.0009	&	0.0432 $\pm$ 0.0041	&	0.0557 $\pm$ 0.0049	&	0.0400 $\pm$ 0.0042	&	0.0172 $\pm$ 0.0045	&	0.0341 $\pm$ 0.0046	\\
\SIII\	&	6312.06	&	0.0152 $\pm$ 0.0022	&	0.0145 $\pm$ 0.0013	&	0.0183 $\pm$ 0.0016	&	0.0148 $\pm$ 0.0008	&	0.0200 $\pm$ 0.0037	&	$\ldots$	&	0.0265 $\pm$ 0.0042	&	0.0219 $\pm$ 0.0049	&	0.0273 $\pm$ 0.0050	\\
\OI\	&	6363.78	&	0.0147 $\pm$ 0.0015	&	0.0143 $\pm$ 0.0023	&	$\ldots$	&	0.0078 $\pm$ 0.0007	&	0.0130 $\pm$ 0.0036	&	0.0195 $\pm$ 0.0043	&	$\ldots$	&	$\ldots$	&	0.0206 $\pm$ 0.0048	\\
\NII\	&	6548.05	&	0.0742 $\pm$ 0.0033	&	0.0332 $\pm$ 0.0023	&	0.0209 $\pm$ 0.0018	&	0.0133 $\pm$ 0.0012	&	0.0221 $\pm$ 0.0045	&	0.0409 $\pm$ 0.0067	&	$\ldots$	&	0.0194 $\pm$ 0.0046	&	0.0296 $\pm$ 0.0065	\\
\HA\	&	6562.82	&	2.8351 $\pm$ 0.0948	&	2.8175 $\pm$ 0.0901	&	2.8170 $\pm$ 0.0874	&	2.7825 $\pm$ 0.0866	&	2.8211 $\pm$ 0.0969	&	2.8167 $\pm$ 0.1077	&	2.7645 $\pm$ 0.0999	&	2.8125 $\pm$ 0.0948	&	2.7920 $\pm$ 0.1000	\\
\NII\	&	6583.46	&	0.2272 $\pm$ 0.0080	&	0.1113 $\pm$ 0.0043	&	0.0643 $\pm$ 0.0028	&	0.0396 $\pm$ 0.0017	&	0.0981 $\pm$ 0.0048	&	0.1247 $\pm$ 0.0081	&	0.0772 $\pm$ 0.0068	&	0.0480 $\pm$ 0.0044	&	0.0695 $\pm$ 0.0080	\\
He I	&	6678.15	&	0.0292 $\pm$ 0.0022	&	0.0286 $\pm$ 0.0019	&	0.0313 $\pm$ 0.0015	&	0.0286 $\pm$ 0.0012	&	0.0242 $\pm$ 0.0040	&	0.0177 $\pm$ 0.0033	&	0.0233 $\pm$ 0.0048	&	0.0210 $\pm$ 0.0029	&	0.0298 $\pm$ 0.0033	\\
\SII\	&	6716.44	&	0.2213 $\pm$ 0.0085	&	0.2733 $\pm$ 0.0100	&	0.1149 $\pm$ 0.0051	&	0.0833 $\pm$ 0.0030	&	0.1954 $\pm$ 0.0087	&	0.3240 $\pm$ 0.0156	&	0.2115 $\pm$ 0.0113	&	0.1044 $\pm$ 0.0063	&	0.2049 $\pm$ 0.0112	\\
\SII\	&	6730.81	&	0.1631 $\pm$ 0.0063	&	0.1854 $\pm$ 0.0077	&	0.0860 $\pm$ 0.0051	&	0.0662 $\pm$ 0.0024	&	0.1353 $\pm$ 0.0070	&	0.2165 $\pm$ 0.0117	&	0.1476 $\pm$ 0.0099	&	0.0777 $\pm$ 0.0054	&	0.1446 $\pm$ 0.0088	\\
 He I	&	7065.20	&	0.0239 $\pm$ 0.0017	&	$\ldots$	&	0.0084 $\pm$ 0.0013	&	0.0329 $\pm$ 0.0020	&	0.0237 $\pm$ 0.0035	&	$\ldots$	&	$\ldots$	&	$\ldots$	&	$\ldots$	\\
\ArIII\	&	7135.79	&	0.0811 $\pm$ 0.0037	&	0.0548 $\pm$ 0.0033	&	0.0926 $\pm$ 0.0062	&	0.0491 $\pm$ 0.0021	&	0.0830 $\pm$ 0.0056	&	0.0623 $\pm$ 0.0091	&	0.0432 $\pm$ 0.0103	&	0.0764 $\pm$ 0.0059	&	0.0729 $\pm$ 0.0116	\\
\OII\	&	7318.99	&	0.0275 $\pm$ 0.0018	&	0.0266 $\pm$ 0.0026	&	0.0194 $\pm$ 0.0035	&	0.0169 $\pm$ 0.0011	&	0.0210 $\pm$ 0.0037	&	0.0345 $\pm$ 0.0096	&	$\ldots$	&	0.0168 $\pm$ 0.0034	&	$\ldots$	\\
\OII\	&	7329.73	&	0.0222 $\pm$ 0.0014	&	0.0117 $\pm$ 0.0019	&	0.0161 $\pm$ 0.0041	&	0.0068 $\pm$ 0.0009	&	$\ldots$	&	$\ldots$	&	$\ldots$	&	$\ldots$	&	$\ldots$	\\
\hline																					
c$_{\text{H}\beta}$	&		&	0.357 $\pm$ 0.041	&	0.233 $\pm$ 0.040	&	0.278 $\pm$ 0.040	&	0.349 $\pm$ 0.040	&	0.045 $\pm$ 0.042	&	0.281 $\pm$ 0.045	&	0.364 $\pm$ 0.043	&	0.251 $\pm$ 0.041	&	0.281 $\pm$ 0.043	\\
EW(\HB)	&		&	34.51 \AA\	&	54.41 \AA\	&	81.85 \AA\	&	130.69 \AA\	&	33.79 \AA\	&	20.21 \AA\	&	36.89 \AA\	&	69.10 \AA\	&	46.38 \AA\	\\
F(\HB)*	&		&	4.2810 $\pm$ 0.0377	&	3.3710 $\pm$ 0.0231	&	6.2220 $\pm$ 0.0203	&	8.0150 $\pm$ 0.0177	&	9.9085 $\pm$ 0.1163	&	2.5849 $\pm$ 0.0401	&	0.9459 $\pm$ 0.0119	&	1.4380 $\pm$ 0.0138	&	0.6082 $\pm$ 0.0077	
\enddata
\tablecomments{Line flux of \HB\ in units of 10$^{-15}$ erg s$^{-1}$ cm$^{-2}$.}
\end{deluxetable*}

\begin{deluxetable*}{cccccccccc} % Table 2c
\tablenum{2}
\label{tab:ratios_allc}
\rotate
\tabletypesize{\scriptsize}
\tablewidth{0pt}
\tablecaption{
Emission-line ratios relative to \HB\ (part 3 of 3).
}
\tablehead{
\colhead{Ion}&\colhead{$\lambda$ [\AA]}&\colhead{\HA\ Dot 145 (K)}&\colhead{\HA\ Dot 151 (K)}&\colhead{\HA\ Dot 157 (K)}&\colhead{\HA\ Dot 173 (K)}&\colhead{\HA\ Dot 174 (K)}&\colhead{\HA\ Dot 194 (K)}&\colhead{\HA\ Dot 218 (K)}&\colhead{\HA\ Dot 303 (K)}
}
\startdata
\OII\	&	3727.43	&	1.8373 $\pm$ 0.0660	&	1.4397 $\pm$ 0.0459	&	1.6679 $\pm$ 0.0916	&	3.1218 $\pm$ 0.1147	&	1.0245 $\pm$ 0.0782	&	3.0036 $\pm$ 0.1413	&	2.1183 $\pm$ 0.0693	&	0.4806 $\pm$ 0.0306	\\
H 10	&	3797.90	&	$\ldots$	&	0.0430 $\pm$ 0.0078	&	$\ldots$	&	$\ldots$	&	$\ldots$	&	$\ldots$	&	$\ldots$	&	$\ldots$	\\
H 9	&	3835.39	&	$\ldots$	&	0.0663 $\pm$ 0.0055	&	$\ldots$	&	$\ldots$	&	$\ldots$	&	$\ldots$	&	$\ldots$	&	0.0707 $\pm$ 0.0142	\\
\NeIII\	&	3868.74	&	0.2877 $\pm$ 0.0173	&	0.2840 $\pm$ 0.0095	&	0.3881 $\pm$ 0.0257	&	0.2530 $\pm$ 0.0162	&	0.4266 $\pm$ 0.0393	&	0.2464 $\pm$ 0.0267	&	0.4695 $\pm$ 0.0189	&	0.0893 $\pm$ 0.0149	\\
He  I + H 8	&	3888.65	&	0.0923 $\pm$ 0.0126	&	0.1838 $\pm$ 0.0069	&	0.1089 $\pm$ 0.0268	&	0.1217 $\pm$ 0.0162	&	0.1299 $\pm$ 0.0348	&	$\ldots$	&	0.1571 $\pm$ 0.0157	&	0.1694 $\pm$ 0.0134	\\
\NeIII\	&	3967.47	&	$\ldots$	&	0.0818 $\pm$ 0.0050	&	$\ldots$	&	$\ldots$	&	0.1075 $\pm$ 0.0262	&	$\ldots$	&	0.0894 $\pm$ 0.0126	&	$\ldots$	\\
\HE\	&	3970.08	&	0.0655 $\pm$ 0.0173	&	0.1489 $\pm$ 0.0057	&	$\ldots$	&	0.0639 $\pm$ 0.0155	&	$\ldots$	&	$\ldots$	&	0.1326 $\pm$ 0.0130	&	0.1423 $\pm$ 0.0108	\\
He I	&	4026.19	&	$\ldots$	&	$\ldots$	&	$\ldots$	&	$\ldots$	&	$\ldots$	&	$\ldots$	&	$\ldots$	&	$\ldots$	\\
\HD\	&	4101.74	&	0.2777 $\pm$ 0.0147	&	0.2645 $\pm$ 0.0073	&	0.2468 $\pm$ 0.0190	&	0.2664 $\pm$ 0.0141	&	0.3085 $\pm$ 0.0304	&	0.2618 $\pm$ 0.0315	&	0.2685 $\pm$ 0.0146	&	0.2638 $\pm$ 0.0119	\\
\HG\	&	4340.47	&	0.4717 $\pm$ 0.0174	&	0.4881 $\pm$ 0.0079	&	0.4914 $\pm$ 0.0267	&	0.4632 $\pm$ 0.0156	&	0.4511 $\pm$ 0.0240	&	0.4698 $\pm$ 0.0267	&	0.4788 $\pm$ 0.0102	&	0.4880 $\pm$ 0.0118	\\
\OIII\	&	4363.21	&	0.0452 $\pm$ 0.0073	&	0.0741 $\pm$ 0.0037	&	0.0934 $\pm$ 0.0140	&	0.0403 $\pm$ 0.0061	&	0.1031 $\pm$ 0.0196	&	0.0872 $\pm$ 0.0166	&	0.0697 $\pm$ 0.0061	&	0.0356 $\pm$ 0.0096	\\
He I	&	4471.48	&	0.0243 $\pm$ 0.0068	&	0.0288 $\pm$ 0.0042	&	$\ldots$	&	$\ldots$	&	$\ldots$	&	$\ldots$	&	0.0300 $\pm$ 0.0055	&	$\ldots$	\\
He II	&	4685.71	&	$\ldots$	&	0.0183 $\pm$ 0.0030	&	$\ldots$	&	$\ldots$	&	$\ldots$	&	$\ldots$	&	$\ldots$	&	$\ldots$	\\
\ArIV\	&	4711.26	&	$\ldots$	&	$\ldots$	&	$\ldots$	&	$\ldots$	&	$\ldots$	&	$\ldots$	&	$\ldots$	&	$\ldots$	\\
\ArIV\	&	4740.12	&	$\ldots$	&	$\ldots$	&	$\ldots$	&	$\ldots$	&	$\ldots$	&	$\ldots$	&	$\ldots$	&	$\ldots$	\\
\HB\	&	4861.33	&	1.0000 $\pm$ 0.0189	&	1.0000 $\pm$ 0.0044	&	1.0000 $\pm$ 0.0286	&	1.0000 $\pm$ 0.0217	&	1.0000 $\pm$ 0.0296	&	1.0000 $\pm$ 0.0347	&	1.0000 $\pm$ 0.0083	&	1.0000 $\pm$ 0.0103	\\
He I	&	4921.93	&	$\ldots$	&	$\ldots$	&	$\ldots$	&	$\ldots$	&	$\ldots$	&	$\ldots$	&	$\ldots$	&	$\ldots$	\\
\OIII\	&	4958.91	&	1.2531 $\pm$ 0.0182	&	1.2114 $\pm$ 0.0050	&	1.4444 $\pm$ 0.0325	&	0.8862 $\pm$ 0.0155	&	1.7178 $\pm$ 0.0467	&	0.8650 $\pm$ 0.0237	&	1.7592 $\pm$ 0.0126	&	0.4433 $\pm$ 0.0071	\\
\OIII\	&	5006.84	&	3.7806 $\pm$ 0.0523	&	3.6408 $\pm$ 0.0161	&	4.3343 $\pm$ 0.0906	&	2.6602 $\pm$ 0.0432	&	5.1173 $\pm$ 0.1105	&	2.6660 $\pm$ 0.0667	&	5.3220 $\pm$ 0.0361	&	1.2795 $\pm$ 0.0163	\\
\NI\	&	5197.90	&	$\ldots$	&	$\ldots$	&	$\ldots$	&	$\ldots$	&	$\ldots$	&	$\ldots$	&	$\ldots$	&	$\ldots$	\\
He I	&	5875.62	&	0.0973 $\pm$ 0.0064	&	0.0983 $\pm$ 0.0034	&	0.0954 $\pm$ 0.0140	&	0.0982 $\pm$ 0.0071	&	0.1138 $\pm$ 0.0172	&	0.1067 $\pm$ 0.0085	&	0.0989 $\pm$ 0.0047	&	0.0957 $\pm$ 0.0065	\\
\OI\	&	6300.30	&	0.0495 $\pm$ 0.0075	&	0.0270 $\pm$ 0.0026	&	$\ldots$	&	0.0652 $\pm$ 0.0042	&	0.0434 $\pm$ 0.0113	&	0.0796 $\pm$ 0.0055	&	0.0400 $\pm$ 0.0033	&	0.0134 $\pm$ 0.0031	\\
\SIII\	&	6312.06	&	$\ldots$	&	0.0129 $\pm$ 0.0024	&	$\ldots$	&	0.0175 $\pm$ 0.0047	&	$\ldots$	&	$\ldots$	&	0.0153 $\pm$ 0.0038	&	$\ldots$	\\
\OI\	&	6363.78	&	$\ldots$	&	0.0077 $\pm$ 0.0015	&	$\ldots$	&	$\ldots$	&	$\ldots$	&	$\ldots$	&	0.0195 $\pm$ 0.0038	&	$\ldots$	\\
\NII\	&	6548.05	&	0.0398 $\pm$ 0.0080	&	0.0200 $\pm$ 0.0032	&	$\ldots$	&	0.0463 $\pm$ 0.0060	&	$\ldots$	&	0.0754 $\pm$ 0.0092	&	0.0228 $\pm$ 0.0039	&	$\ldots$	\\
\HA\	&	6562.82	&	2.8227 $\pm$ 0.1015	&	2.7869 $\pm$ 0.0852	&	2.7835 $\pm$ 0.1171	&	2.8077 $\pm$ 0.1048	&	2.7875 $\pm$ 0.1229	&	2.7511 $\pm$ 0.1288	&	2.8167 $\pm$ 0.0907	&	2.7664 $\pm$ 0.0881	\\
\NII\	&	6583.46	&	0.0998 $\pm$ 0.0079	&	0.0534 $\pm$ 0.0023	&	0.0480 $\pm$ 0.0084	&	0.1594 $\pm$ 0.0079	&	0.0175 $\pm$ 0.0091	&	0.2462 $\pm$ 0.0153	&	0.0716 $\pm$ 0.0043	&	0.0154 $\pm$ 0.0037	\\
He I	&	6678.15	&	0.0177 $\pm$ 0.0040	&	0.0249 $\pm$ 0.0017	&	0.0259 $\pm$ 0.0063	&	0.0171 $\pm$ 0.0033	&	0.0588 $\pm$ 0.0094	&	0.0325 $\pm$ 0.0084	&	0.0311 $\pm$ 0.0059	&	0.0193 $\pm$ 0.0029	\\
\SII\	&	6716.44	&	0.2463 $\pm$ 0.0124	&	0.1663 $\pm$ 0.0061	&	0.1779 $\pm$ 0.0111	&	0.3238 $\pm$ 0.0144	&	0.1849 $\pm$ 0.0195	&	0.4213 $\pm$ 0.0220	&	0.1882 $\pm$ 0.0104	&	0.0304 $\pm$ 0.0046	\\
\SII\	&	6730.81	&	0.1774 $\pm$ 0.0140	&	0.1090 $\pm$ 0.0046	&	0.1270 $\pm$ 0.0110	&	0.2113 $\pm$ 0.0117	&	0.1010 $\pm$ 0.0161	&	0.3141 $\pm$ 0.0169	&	0.1128 $\pm$ 0.0058	&	0.0230 $\pm$ 0.0045	\\
 He I	&	7065.20	&	0.0190 $\pm$ 0.0053	&	0.0201 $\pm$ 0.0018	&	$\ldots$	&	$\ldots$	&	$\ldots$	&	$\ldots$	&	$\ldots$	&	0.0225 $\pm$ 0.0039	\\
\ArIII\	&	7135.79	&	0.0721 $\pm$ 0.0086	&	0.0555 $\pm$ 0.0033	&	0.0581 $\pm$ 0.0097	&	$\ldots$	&	0.0714 $\pm$ 0.0171	&	0.0661 $\pm$ 0.0133	&	0.0781 $\pm$ 0.0085	&	$\ldots$	\\
\OII\	&	7318.99	&	0.0288 $\pm$ 0.0070	&	0.0197 $\pm$ 0.0027	&	$\ldots$	&	0.0304 $\pm$ 0.0045	&	$\ldots$	&	0.0342 $\pm$ 0.0048	&	0.0194 $\pm$ 0.0033	&	$\ldots$	\\
\OII\	&	7329.73	&	$\ldots$	&	0.0173 $\pm$ 0.0024	&	$\ldots$	&	$\ldots$	&	$\ldots$	&	$\ldots$	&	0.0108 $\pm$ 0.0023	&	0.0416 $\pm$ 0.0047	\\
\hline																			
c$_{\text{H}\beta}$	&		&	0.009 $\pm$ 0.043	&	0.035 $\pm$ 0.039	&	0.131 $\pm$ 0.047	&	0.168 $\pm$ 0.044	&	0.162 $\pm$ 0.049	&	0.244 $\pm$ 0.051	&	0.363 $\pm$ 0.040	&	0.002 $\pm$ 0.040	\\
EW(\HB)	&		&	19.54 \AA\	&	90.98 \AA\	&	99.07 \AA\	&	21.98 \AA\	&	22.45 \AA\	&	11.98 \AA\	&	44.26 \AA\	&	65.37 \AA\	\\
F(\HB)*	&		&	2.2120 $\pm$ 0.0295	&	3.5045 $\pm$ 0.0108	&	0.7019 $\pm$ 0.0142	&	1.3448 $\pm$ 0.0206	&	0.5725 $\pm$ 0.0120	&	1.4358 $\pm$ 0.0353	&	2.8074 $\pm$ 0.0165	&	1.9084 $\pm$ 0.0139	
\enddata
\tablecomments{Line flux of \HB\ in units of 10$^{-15}$ erg s$^{-1}$ cm$^{-2}$.}
\end{deluxetable*}

\section{Abundance Analysis} % Section 3.
\label{sec:analysis}

\subsection{Line Ratios and Diagnostic Diagram} % Section 3.1.
\label{sec:line_ratios}

\begin{figure*} % Figure 3
\figurenum{3}
\plotone{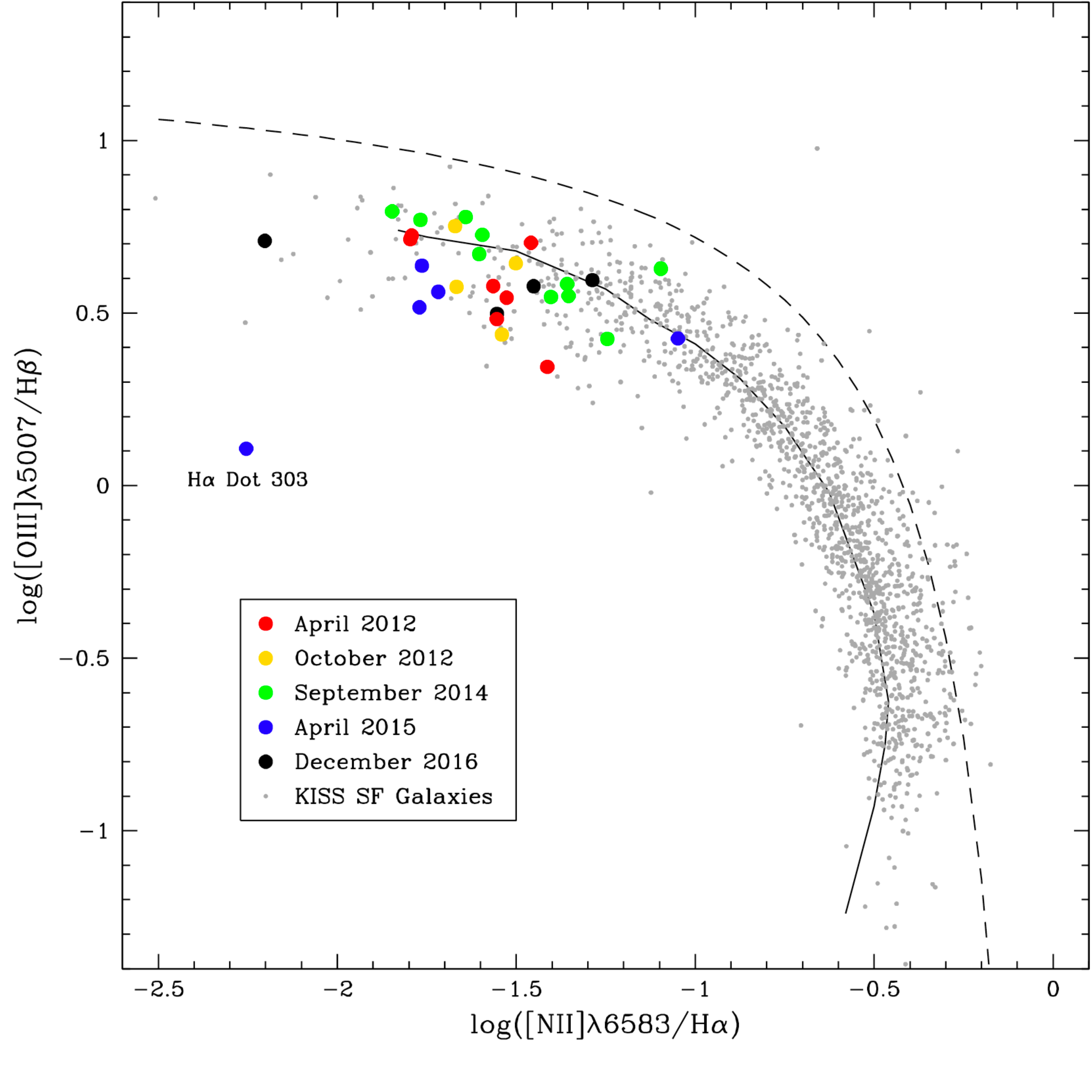}
\caption{
Spectral activity diagnostic diagram of \HA\ Dots with direct-method abundances.
Color-coding specifies the observing run:
Red points from April 2012, gold points from October 2012, green points from September 2015, blue points from April 2015, and black points from December 2016.
The small gray points are star-forming galaxies from the KPNO International Spectroscopic Survey (KISS; \citealp{bib:Salzer2000, bib:Salzer2001}), and are included to provide context.
The solid line represents a sequence of nebular models from \citet{bib:DopitaEvans1986}, while the dashed line represents an empirically defined demarcation between starburst galaxies and AGN \citep{bib:Kauffmann2003}.
The line ratios of \HA\ Dots are consistent with low-abundance, high-excitation star-forming systems.
The single blue point well-separated from the other sources is labeled as \HA\ Dot 303.
Also known as Leoncino \citep{bib:Hirschauer2016, bib:McQuinn2020, bib:Aver2021}, this galaxy is also referenced as \Alecxy\ in the Survey of \HI\ in Extremely Low-mass Dwarfs (SHIELD) sample \citet{bib:Cannon2011}.
It is among the most extremely metal-poor (XMP) star-forming galaxies known ($\sim$2\% solar).
}
\label{fig:DDDots}
\end{figure*}

\indent Results of the emission-line measurement and analysis, presented as reddening-corrected line ratios relative to \HB, are listed in Table \ref{tab:ratios_alla}.
In addition, we include the measured values of \cHB\ (see \S \ref{sec:datareduction} for details), the EW of the \HB\ emission line, and the line flux of the \HB\ emission line in units of 10$^{-15}$ erg s$^{-1}$ cm$^{-2}$.
The spectroscopic instrument used for the observation is noted for each \HA\ Dot.
%In the cases where a source was observed twice, the observations are treated as separate and both sets of values are included here.
We note that the \OII$\lambda$3727 line listed here is a blend of the unresolved \OII$\lambda\lambda$3726,3729 doublet.
In cases where a source was observed twice, only the higher-quality data employed for subsequent abundance analyses are included.
\\
\indent Locations of the \HA\ Dots on a standard emission-line ratio diagnostic diagram (e.g., \citealp{bib:Baldwin1981, bib:VeilleuxOsterbrock1987}) are presented in Figure \ref{fig:DDDots}.
Sources are color coded by observing run; red points are from April 2012, gold points from October 2012, green points from September 2014, blue points from April 2015, and black points from December 2016.
\HA\ Dots are plotted over star-forming galaxies from the KPNO International Spectroscopic Survey (KISS; \citealp{bib:Salzer2000, bib:Salzer2001}) as gray dots, which constitutes a comparison sample of star-forming systems spanning a wide range in metallicity.
The solid line represents a sequence of nebular models from \citet{bib:DopitaEvans1986}, while the dashed line represents an empirically-defined demarcation between starburst galaxies and AGNs from \citet{bib:Kauffmann2003}.
\\
\indent The locations of the \HA\ Dots on this plot signify that they are all consistent with fairly high-excitation, low-abundance systems.
A single labeled blue point, which is well-separated from the other sources, represents \HA\ Dot 303.
This object was originally included in the ALFALFA-based Survey of \HI\ in Extremely Low-mass Dwarfs (SHIELD) sample \citep{bib:Cannon2011}, designated as \Alecxy.
It was subsequently identified as an \HA\ Dot during analysis of its narrowband imaging data.
%Narrowband \HA\ imaging data of this galaxy was then run through the \HA\ Dots processing software, which identified it as an \HA\ Dot candidate.
Spectral followup confirmed it as a \emph{bona fide} local, dwarf star-forming system.
With an abundance of $\sim$2\% solar (assuming solar metallicity of \abun\ = 8.69; \citealp{bib:Asplund2009}), this galaxy (Leoncino; \citealp{bib:Hirschauer2016, bib:McQuinn2020, bib:Aver2021}) is among the most extremely metal-poor (XMP) systems known.

\subsection{Determination of Density and Temperature} % Section 3.2.
\label{sec:tempdens}

\indent Calculations of both the density and temperature of the electron gas were carried out using the Emission Line Spectrum Analyzer (\texttt{ELSA}) program \citep{bib:Johnson2006}.
The electron density (\Ne) is determined via the \SII$\lambda$6716/\SII$\lambda$6731 line ratio, where the emission originates from similar energy levels of the same ion, but the transitions have distinctly different critical densities \citep{bib:OsterbrockFerland2006}.
Because the relative populations of the two levels depend on the \Ne, the ratio of intensities will follow suit.
In the majority of measurable cases, the \Ne\ is roughly 100 e$^{-}$ cm$^{-3}$.
For galaxies in which the value of \Ne\ was not derived from our observations, we assume a density of 100 e$^{-}$ cm$^{-3}$.
Our abundance analysis assumes that the density does not vary across ionization zones.
\\
\indent The temperature of the electron gas (\Te) was determined by the oxygen line ratio \OIII$\lambda$4363/\OIII$\lambda\lambda$4959,5007, which exhibits a strong temperature dependence because the emission arises from two widely separated energy levels in the same ionic species (e.g., \citealp{bib:OsterbrockFerland2006}).
Metallicities derived based on the measurement of the \Te\ are commonly referred to as ``direct-method" abundances, and are considered more robust than techniques which rely on, for example, empirical calibrations utilizing strong emission line (SEL) ratios.
From our observations, the values of \Te\ for the \HA\ Dot sample range from 11570 $\pm$ 354 K to 19760 $\pm$ 2267 K, with an average temperature of $\sim$15000 K.
For most \HA\ Dots, uncertainties in the \Te\ value are relatively small (a $\sim$few hundred K).
In a handful of cases, however, the uncertainties were found to be quite large ($\gtrsim$ 2000 K).
These subsequently manifest as large error values in the \Te-method abundance determinations.
\HA\ Dots exhibit relatively high temperatures as compared with an \HII\ region of typical abundance (\Te\ $\approx$ 6000 -- 10000 K; \citealp{bib:OsterbrockFerland2006}), consistent with expectation based on their locations in the line ratio diagnostic diagram discussed in \S \ref{sec:line_ratios}.

\subsection{Determination of Elemental Abundances} % Section 3.3.
\label{sec:ICFs}

\indent We use the program \texttt{ELSA} \citep{bib:Johnson2006} to calculate ionic abundances relative to hydrogen.
The temperature for the low-ionization O$^{+}$ zone was estimated by using the algorithm presented in \citet{bib:Skillman1994}, derived by \citet{bib:Pagel1992}, and based on the nebular models of \citet{bib:Stasinska1990},
\[
t_{e}(\text{O}^{+}) = 2[t_{e}(\text{O}^{++})^{-1} + 0.8]^{-1},
\]
\noindent where $t_{e}$ are temperatures measured in units of 10$^{4}$ K.
Following the standard practice, the total oxygen abundance is assumed to be given by
\[
\frac{\text{O}}{\text{H}} = \frac{\text{O}^{+}}{\text{H}^{+}} + \frac{\text{O}^{++}}{\text{H}^{+}},
\]
and oxygen abundance is used as a proxy for the overall metallicity of a system, because it is the most prevalent heavy element and all relevant ionization states (O$^{+}$ and O$^{++}$) are observable and strongly emitting.
Additional ionization states for other elements that are present in the nebula but do not emit in the optical region of the spectrum are accounted for with ionization correction factors (ICFs).
We use the prescriptions given by \citet{bib:PeimbertCostero1969} for the ICFs for N and Ne:
\[
\text{ICF(N)} = \frac{\text{N}}{\text{N}^{+}} = \frac{\text{O}}{\text{O}^{+}},
\]
\[
\text{ICF(Ne)} = \frac{\text{Ne}}{\text{Ne}^{++}} = \frac{\text{O}}{\text{O}^{++}}.
\]
ICFs for S and Ar are adopted from \citet{bib:Izotov1994}:
\[
\text{ICF(S)} = \frac{\text{S}}{\text{S}^{+} + \text{S}^{++}} = 
\]
\[
(0.013 + x\{5.10 + x[-12.78 + x(14.77 - 6.11x)]\})^{-1},
\]
\[
\text{ICF(Ar}) = \frac{\text{Ar}}{\text{Ar}^{++}} = [0.15 + x(2.39 - 2.64x)]^{-1},
\]
\[
\text{where}\ x = \frac{\text{O}^{+}}{\text{O}}.
\]
\citet{bib:Izotov1994} suggest that the flux for O$^{+++}$ can be inferred by
\[
\frac{\text{O}^{+++}}{\text{O}^{++}} = \frac{\text{He}^{++}}{\text{He}^{+}}.
\]
In the few cases in which we observe He$^{++}$ in the spectra, however, the line is fairly noisy, and we assume that the amount of O$^{+++}$ in the nebulae is negligible.
Using the ionic abundances and ionization correction factors given above, we calculate the 
He/H, O/H, N/H, N/O, Ne/H, Ne/O, S/H, S/O, Ar/H, and Ar/O ratios for each data set.  
These values are presented in Table \ref{tab:abundances_all}.
Recent work by \citet{bib:Amayo2021} presented issues related to oxygen abundance and degree of ionization on the efficacy of ICFs, which may contribute considerably to elemental abundance uncertainties.
%\textbf{In addition, \citet{bib:JuandeDiosRodriguez2017} analyzed the impact of atomic data (transition probabilities and collision strengths) on ionic abundance determinations, finding some biases which result in value discrepancies.
%We have found that the ionic data in question and the}
The extremity required of such parameters to become substantially manifest is not exhibited by the \HA\ Dots sample, therefore there is no indication of such biases.
% \textbf{(elevated ionization and high electron density, respectively)} 
Furthermore, we emphasize that our helium abundances are only approximate, as they do not account for underlying absorption or neutral helium, which may be non-negligible in these sources.
%Because the \texttt{ELSA} code was originally developed for the study of planetary nebulae (PNe) abundances, then only later adopted for \HII\ regions, neutral helium is not considered.}
\\
\indent We note that abundance values presented in this study for \HA\ Dot 303 are based on the KPNO 4m telescope KOSMOS spectrum in order to keep its results consistent with observational data of the other \HA\ Dots.
In \citet{bib:Hirschauer2016}, deeper observations with the MMT 6.5m telescope provided higher-quality spectral data, with an oxygen abundance found to be \abun\ = 7.02 $\pm$ 0.03.
More recently, \citet{bib:Aver2021} has updated the metallicity to \abun\ = 7.06 $\pm$ 0.03 based on observations from the Large Binocular Telescope (LBT).

\subsection{Estimations of H$\alpha$ Dot Stellar Mass and Star-Formation Rate} % Section 3.4.
\label{sec:MassAndSFR}

\begin{figure*} % Figure 4
\figurenum{4}
\plotone{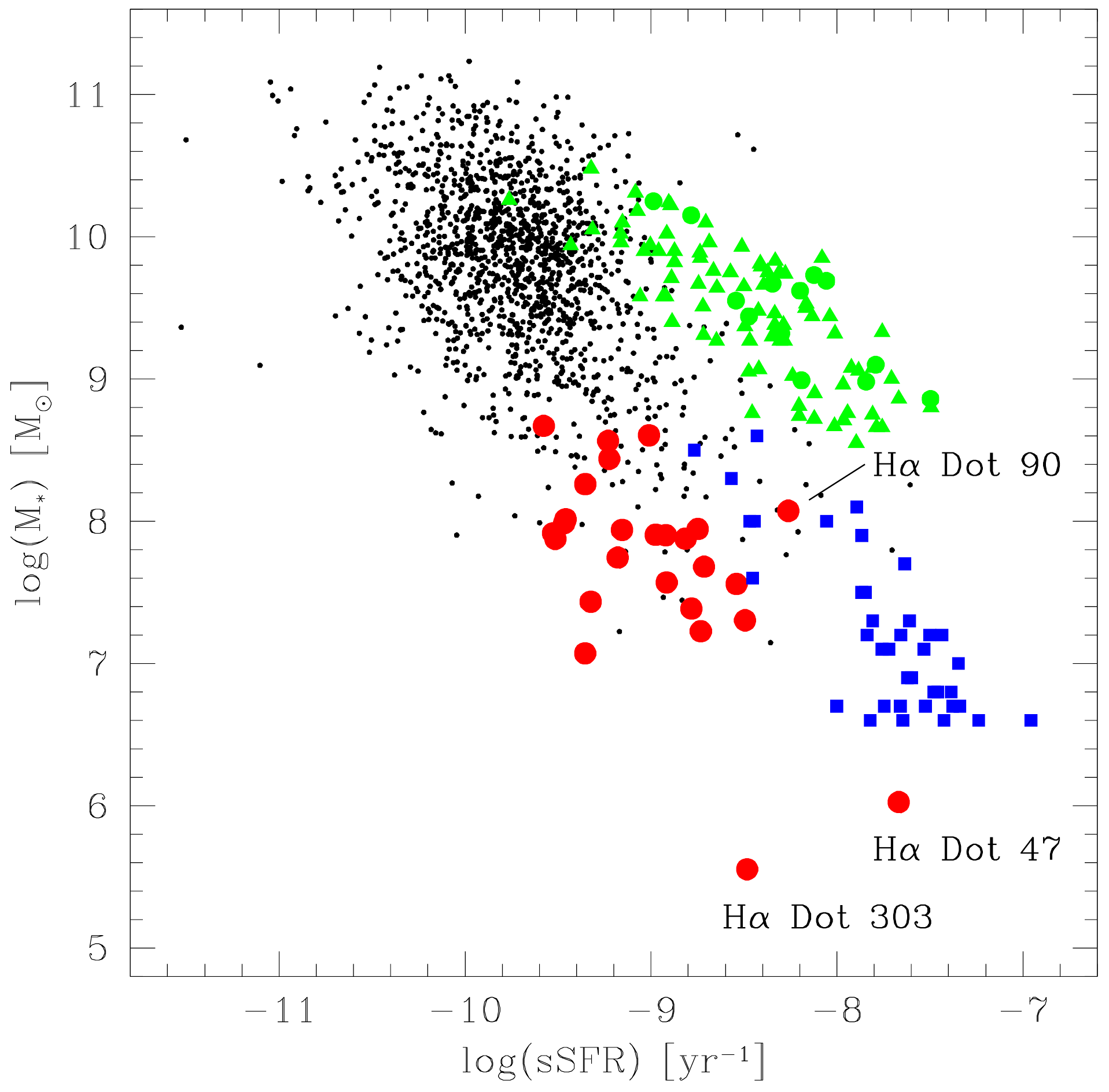}
\caption{
Comparison of the specific star-formation rate (sSFR) of the \HA\ Dots sample (red dots) with green peas (green triangles; \citealp{bib:Cardamone2009}, and green dots; \citealp{bib:Brunker2020}), blueberries (blue squares; \citealp{bib:Yang2017}), and \HA-selected KISS star-forming galaxies (black points) as a function of stellar mass.
The sSFR of \HA\ Dots is typically elevated as compared to the KISS sample, but is consistent with the lower range of green peas.
%The blueberries lie between the \HA\ Dots and the green peas, although \HA\ Dot 90 has properties similar to the blueberries.
The blueberries lie between the \HA\ Dots and the green peas, although two of them (\HA\ Dot 90 and \HA\ Dot 47) have properties similar to the blueberries.
We also label the low mass, low metallicity \HA\ Dot 303 (Leoncino).
%noting that the properties of some \HA\ Dots (labeled) demonstrate overlap with the blueberries.}.
Average \HA\ Dot stellar mass, is $\sim$1.5 dex lower compared to green peas, and roughly consistent (though slightly higher on average) with the range of the blueberries.
}
\label{fig:sSFR}
\end{figure*}

\indent Stellar mass estimates of the \HA\ Dots sample were obtained by performing SED model fitting based on available photometric data for each object.
We utilized the Code Investigating Galaxy Emission (\texttt{CIGALE}) software \citep{bib:Noll2009} to carry out the SED fitting, following the procedures described in \citet{bib:Janowiecki2017} and \citet{bib:Hirschauer2018}.
Photometric data from a variety of sources was used, including UV fluxes from the \emph{Galaxy Evolution Explorer} (\emph{GALEX}; \citealp{bib:Martin2005, bib:Morrissey2007}), optical photometry from SDSS Data Release 12 (DR12; \citealp{bib:Alam2015}) and the \HA\ Dots catalogs \citep{bib:Kellar2012, bib:Salzer2020}, and IR photometry including near-IR from the \emph{Two Micron All Sky Survey} (\emph{2MASS}; \citealp{bib:Skrutskie2006}) and mid-IR from the \emph{Wide-field Infrared Survey Explorer} (\emph{WISE}; \citealp{bib:Wright2010}) databases.
Despite the relatively faint nature of the objects in our sample, the majority of the \HA\ Dots yielded SEDs that included flux points from the UV to the IR, resulting in robust stellar mass determinations.
\\
\indent The star-formation rate (SFR) values of the \HA\ Dots sample were taken directly from the survey papers \citep{bib:Kellar2012, bib:Salzer2020}.
The SFRs are derived using the distances listed in Table \ref{tab:Dot_info} and the \HA\ fluxes measured from the original survey images.
We adopt the \citet{bib:Kennicutt1998} relation with a \citet{bib:Salpeter1955} initial mass function (IMF) to convert from \HA\ luminosity to SFR:
\[
%\mathrm{SFR} = \frac{L_{\mathrm{H\alpha}}}{7.9 \times 10^{42}},
\mathrm{SFR} = 7.9 \times 10^{-42} \cdot L_{\mathrm{H\alpha}},
\]
where SFR is in M$_{\odot}$ yr$^{-1}$ and $L_{\mathrm{H\alpha}}$ is the \HA\ luminosity in erg s$^{-1}$.
Uncertainties in the SFR values are dominated by uncertainties in distance measurements, with typical values of 5-10\%.
Typical uncertainties of the SFR are therefore of order 10\%.
\\
\indent Finally, the SFR measured as a function of the stellar mass (the specific star-formation rate; sSFR) was estimated for each object based on the two aforementioned parameters.
Uncertainties in the sSFR are dominated by errors in the stellar mass estimates, which are typically 20-30\% \citep{bib:Janowiecki2017, bib:Hirschauer2018}.
Typical uncertainties in the sSFR are therefore also of order 20-30\%.
We find that the \HA\ Dots sample have elevated sSFR values typical of dwarf star-forming systems, but are not as extreme as other types of objects such as ``green peas" originally discovered in SDSS \citep{bib:Cardamone2009}.
For our twenty-six sources, values of log(sSFR) range from --9.58 to --7.76, with an average of --8.98.
These span a range approximately equivalent to that of green peas \citep{bib:Cardamone2009}, luminous compact galaxies in the SDSS (LCGs; \citealp{bib:Izotov2011}), \emph{VIMOS Ultra Deep Survey} (\emph{VUDS}) star-forming dwarf galaxies (SFDGs; \citealp{bib:Calabro2017}), and \OIII-detected star-forming galaxies selected from KISS \citep{bib:Brunker2020}.
The sSFR range for \HA\ Dots is generally higher than that of the \HA-selected KISS galaxy sample \citep{bib:Salzer2000, bib:Salzer2001} and the neutral gas detected XMP Leo P \citep{bib:McQuinn2015a}, but lower than well-known XMPs such as I Zw 18 \citep{bib:Annibali2013}, SBS 0335-052W \citep{bib:Schneider2016}, and DDO 68 \citep{bib:Sacchi2016}.
\\
\indent The stellar mass, SFR, and sSFR estimates of the \HA\ Dots from this sample are summarized in Table \ref{tab:Dot_info}.
A plot comparing the sSFR and stellar mass estimates for \HA\ Dots and three other samples of star-forming dwarf systems is presented as Figure \ref{fig:sSFR}.
\HA\ Dots are plotted as red filled circles, green pea systems are presented as green points (triangles for \citealp{bib:Cardamone2009}, dots for \OIII-selected KISS galaxies identified by \citealp{bib:Brunker2020}), and blueberries \citep{bib:Yang2017} are illustrated as blue squares.
\HA-selected star-forming galaxies representing low-$z$ KISS sources are included as black points.
The \HA\ Dots sample overlaps the low-mass portion of the KISS star-forming galaxies, but extends to lower masses (e.g., two \HA\ Dots with stellar masses near or below 10$^{6}$ M$_{\odot}$).
%Some of these lower-mass galaxies exhibit extreme sSFR values, rivaling some of the green pea galaxies.
%The set of the most extreme \HA\ Dots overlaps with the properties of the \citet{bib:Yang2017} blueberries.
The majority of \HA\ Dots bifurcate the mass range of the green peas and blueberries.
Some of these lower-mass \HA\ Dots exhibit extreme sSFR, however the bulk remain roughly consistent with the lower range of green peas, and only show slight overlap with the higher-sSFR blueberries.
In general, the locations of the \HA\ Dots, blueberries, and green peas in Figure \ref{fig:sSFR} suggests that they represent a rough continuum of star-forming galaxies in sSFR space.
The sole \HA\ Dot located within the area occupied by the blueberries is \HA\ Dot 90.
It is the most extreme system in the current sample, and exhibits characteristics of the blueberries despite being very nearby ($\sim$70 Mpc).
In addition, \HA\ Dot 47 shows similarities in sSFR with the blueberries sample, but possesses a lower stellar mass.
%Each of these are labeled in Figure \ref{fig:sSFR} for reference.}

\begin{deluxetable*}{ccccccccccccc} % Table 3
\tablenum{3}
\label{tab:abundances_all}
\rotate
\tabletypesize{\tiny}
\tablewidth{0pt}
\tablecaption{
Electron Temperatures, Densities, and Elemental Abundances.
}
\tablehead{
\colhead{\HA\ Dot}&\colhead{O~{\textsc i}{\textsc i}{\textsc i} $t_{e}$}&\colhead{S~{\textsc i}{\textsc i} $n_{e}$}&\colhead{12+log(He/H)}&\colhead{12+log(N/H)}&\colhead{log(N/O)}&\colhead{12+log(O/H)}&\colhead{12+log(Ne/H)}&\colhead{log(Ne/O)}&\colhead{12+log(S/H)}&\colhead{log(S/O)}&\colhead{12+log(Ar/H)}&\colhead{log(Ar/O)}
\\
\colhead{(Instrument)}&\colhead{[K]}&\colhead{[cm$^{-3}$]}&\colhead{}&\colhead{}&\colhead{}&\colhead{}&\colhead{}&\colhead{}&\colhead{}&\colhead{}&\colhead{}&\colhead{}
}
\startdata
2 (K)	&	17010 $\pm$ 627	&	58.0 $\pm$ 47.8	&	10.882 $\pm$ 0.018	&	5.952 $\pm$ 0.044	&	-1.818 $\pm$ 0.037	&	7.771 $\pm$ 0.026	&	7.079 $\pm$ 0.034	&	-0.693 $\pm$ 0.016	&	5.575 $\pm$ 0.020	&	-2.196 $\pm$ 0.013	&	5.158 $\pm$ 0.044	&	-2.613 $\pm$ 0.036	\\
4 (RC)	&	14240 $\pm$ 473	&	100 $\pm$ ---	&	10.840 $\pm$ 0.020	&	6.342 $\pm$ 0.052	&	-1.529 $\pm$ 0.040	&	7.872 $\pm$ 0.041	&	7.121 $\pm$ 0.047	&	-0.752 $\pm$ 0.012	&	6.225 $\pm$ 0.078	&	-1.648 $\pm$ 0.060	&	5.505 $\pm$ 0.038	&	-2.367 $\pm$ 0.025	\\
12 (RC)	&	15990 $\pm$ 4185	&	17.8 $\pm$ 86.7	&	10.873 $\pm$ 0.049	&	6.338 $\pm$ 0.280	&	-1.279 $\pm$ 0.091	&	7.617 $\pm$ 0.342	&	6.867 $\pm$ 0.412	&	-0.750 $\pm$ 0.055	&	5.516 $\pm$ 0.218	&	-2.101 $\pm$ 0.088	&	5.314 $\pm$ 0.274	&	-2.302 $\pm$ 0.111	\\
20 (RC)	&	18370 $\pm$ 2086	&	100 $\pm$ ---	&	10.831 $\pm$ 0.070	&	6.021 $\pm$ 0.100	&	-1.472 $\pm$ 0.079	&	7.494 $\pm$ 0.084	&	6.692 $\pm$ 0.111	&	-0.804 $\pm$ 0.056	&	5.468 $\pm$ 0.058	&	-2.027 $\pm$ 0.035	&	5.176 $\pm$ 0.124	&	-2.320 $\pm$ 0.108	\\
31 (RC)	&	14470 $\pm$ 926	&	306 $\pm$ 250	&	10.998 $\pm$ 0.047	&	6.350 $\pm$ 0.130	&	-1.504 $\pm$ 0.113	&	7.855 $\pm$ 0.080	&	7.037 $\pm$ 0.093	&	-0.818 $\pm$ 0.025	&	5.458 $\pm$ 0.068	&	-2.397 $\pm$ 0.045	&	5.537 $\pm$ 0.083	&	-2.318 $\pm$ 0.064	\\
34 (RC)	&	16780 $\pm$ 1219	&	100 $\pm$ ---	&	10.702 $\pm$ 0.097	&	6.137 $\pm$ 0.073	&	-1.413 $\pm$ 0.050	&	7.550 $\pm$ 0.070	&	6.798 $\pm$ 0.146	&	-0.752 $\pm$ 0.120	&	5.667 $\pm$ 0.054	&	-1.883 $\pm$ 0.029	&	--- $\pm$ ---	&	--- $\pm$ ---	\\
40 (RC)	&	14810 $\pm$ 1445	&	241 $\pm$ 259	&	10.745 $\pm$ 0.069	&	6.371 $\pm$ 0.125	&	-1.381 $\pm$ 0.080	&	7.753 $\pm$ 0.118	&	6.932 $\pm$ 0.139	&	-0.821 $\pm$ 0.040	&	5.438 $\pm$ 0.091	&	-2.316 $\pm$ 0.052	&	--- $\pm$ ---	&	--- $\pm$ ---	\\
43 (K)	&	12970 $\pm$ 480	&	24.4 $\pm$ 56.6	&	10.909 $\pm$ 0.027	&	6.568 $\pm$ 0.042	&	-1.401 $\pm$ 0.024	&	7.969 $\pm$ 0.046	&	7.308 $\pm$ 0.056	&	-0.662 $\pm$ 0.020	&	6.398 $\pm$ 0.097	&	-1.572 $\pm$ 0.079	&	5.675 $\pm$ 0.052	&	-2.294 $\pm$ 0.038	\\
47 (RC)	&	17420 $\pm$ 2221	&	100 $\pm$ ---	&	--- $\pm$ ---	&	6.068 $\pm$ 0.174	&	-1.523 $\pm$ 0.150	&	7.590 $\pm$ 0.119	&	6.815 $\pm$ 0.150	&	-0.775 $\pm$ 0.060	&	5.137 $\pm$ 0.095	&	-2.452 $\pm$ 0.071	&	--- $\pm$ ---	&	--- $\pm$ ---	\\
53 (K)	&	11570 $\pm$ 354	&	61.5 $\pm$ 20.6	&	10.933 $\pm$ 0.015	&	6.948 $\pm$ 0.031	&	-1.199 $\pm$ 0.012	&	8.146 $\pm$ 0.039	&	7.489 $\pm$ 0.047	&	-0.658 $\pm$ 0.013	&	6.496 $\pm$ 0.068	&	-1.652 $\pm$ 0.049	&	5.738 $\pm$ 0.032	&	-2.409 $\pm$ 0.013	\\
79 (K)	&	12710 $\pm$ 466	&	100 $\pm$ ---	&	10.901 $\pm$ 0.013	&	6.438 $\pm$ 0.037	&	-1.575 $\pm$ 0.014	&	8.013 $\pm$ 0.044	&	7.356 $\pm$ 0.052	&	-0.658 $\pm$ 0.010	&	6.350 $\pm$ 0.059	&	-1.662 $\pm$ 0.028	&	5.477 $\pm$ 0.042	&	-2.536 $\pm$ 0.022	\\
81 (K)	&	12750 $\pm$ 137	&	73.9 $\pm$ 73.8	&	10.907 $\pm$ 0.015	&	6.592 $\pm$ 0.019	&	-1.484 $\pm$ 0.017	&	8.076 $\pm$ 0.014	&	7.398 $\pm$ 0.018	&	-0.680 $\pm$ 0.008	&	6.403 $\pm$ 0.036	&	-1.674 $\pm$ 0.032	&	5.792 $\pm$ 0.026	&	-2.285 $\pm$ 0.025	\\
90 (K)	&	15930 $\pm$ 114	&	156 $\pm$ 22.9	&	10.921 $\pm$ 0.007	&	6.196 $\pm$ 0.016	&	-1.684 $\pm$ 0.016	&	7.880 $\pm$ 0.007	&	7.164 $\pm$ 0.009	&	-0.717 $\pm$ 0.004	&	6.000 $\pm$ 0.018	&	-1.879 $\pm$ 0.016	&	5.223 $\pm$ 0.011	&	-2.658 $\pm$ 0.009	\\
116 (RC)	&	12490 $\pm$ 646	&	100 $\pm$ ---	&	10.939 $\pm$ 0.017	&	6.577 $\pm$ 0.058	&	-1.585 $\pm$ 0.019	&	8.161 $\pm$ 0.053	&	7.468 $\pm$ 0.067	&	-0.695 $\pm$ 0.025	&	6.483 $\pm$ 0.101	&	-1.680 $\pm$ 0.069	&	5.679 $\pm$ 0.056	&	-2.484 $\pm$ 0.024	\\
124 (K)	&	12770 $\pm$ 844	&	100 $\pm$ ---	&	10.907 $\pm$ 0.032	&	6.481 $\pm$ 0.070	&	-1.538 $\pm$ 0.032	&	8.021 $\pm$ 0.080	&	7.371 $\pm$ 0.094	&	-0.650 $\pm$ 0.021	&	5.884 $\pm$ 0.055	&	-2.137 $\pm$ 0.029	&	5.530 $\pm$ 0.090	&	-2.491 $\pm$ 0.064	\\
127 (K)	&	18200 $\pm$ 1426	&	100 $\pm$ ---	&	10.796 $\pm$ 0.040	&	6.057 $\pm$ 0.060	&	-1.471 $\pm$ 0.040	&	7.528 $\pm$ 0.060	&	6.908 $\pm$ 0.072	&	-0.620 $\pm$ 0.021	&	6.076 $\pm$ 0.091	&	-1.451 $\pm$ 0.055	&	5.072 $\pm$ 0.114	&	-2.456 $\pm$ 0.105	\\
131 (K)	&	13120 $\pm$ 398	&	62.1 $\pm$ 98.3	&	10.838 $\pm$ 0.026	&	6.540 $\pm$ 0.057	&	-1.476 $\pm$ 0.048	&	8.017 $\pm$ 0.039	&	7.188 $\pm$ 0.047	&	-0.830 $\pm$ 0.017	&	6.431 $\pm$ 0.101	&	-1.585 $\pm$ 0.089	&	5.661 $\pm$ 0.041	&	-2.357 $\pm$ 0.032	\\
138 (K)	&	14870 $\pm$ 1430	&	100 $\pm$ ---	&	10.859 $\pm$ 0.055	&	6.336 $\pm$ 0.108	&	-1.498 $\pm$ 0.054	&	7.833 $\pm$ 0.116	&	7.158 $\pm$ 0.132	&	-0.674 $\pm$ 0.024	&	6.338 $\pm$ 0.156	&	-1.493 $\pm$ 0.069	&	5.455 $\pm$ 0.115	&	-2.378 $\pm$ 0.072	\\
145 (K)	&	12380 $\pm$ 758	&	30.3 $\pm$ 91.5	&	10.888 $\pm$ 0.027	&	6.614 $\pm$ 0.074	&	-1.373 $\pm$ 0.043	&	7.986 $\pm$ 0.078	&	7.274 $\pm$ 0.093	&	-0.710 $\pm$ 0.026	&	5.814 $\pm$ 0.055	&	-2.172 $\pm$ 0.034	&	5.625 $\pm$ 0.079	&	-2.361 $\pm$ 0.052	\\
151 (K)	&	15440 $\pm$ 352	&	100 $\pm$ ---	&	10.888 $\pm$ 0.012	&	6.149 $\pm$ 0.031	&	-1.674 $\pm$ 0.024	&	7.822 $\pm$ 0.020	&	7.079 $\pm$ 0.025	&	-0.742 $\pm$ 0.010	&	6.049 $\pm$ 0.057	&	-1.772 $\pm$ 0.051	&	5.308 $\pm$ 0.029	&	-2.514 $\pm$ 0.021	\\
157 (K)	&	15850 $\pm$ 1137	&	100 $\pm$ ---	&	10.863 $\pm$ 0.064	&	6.111 $\pm$ 0.101	&	-1.607 $\pm$ 0.079	&	7.718 $\pm$ 0.080	&	7.029 $\pm$ 0.094	&	-0.688 $\pm$ 0.027	&	5.494 $\pm$ 0.058	&	-2.224 $\pm$ 0.032	&	5.303 $\pm$ 0.094	&	-2.415 $\pm$ 0.072	\\
173 (K)	&	13570 $\pm$ 843	&	100 $\pm$ ---	&	10.885 $\pm$ 0.030	&	6.465 $\pm$ 0.062	&	-1.412 $\pm$ 0.024	&	7.878 $\pm$ 0.073	&	7.246 $\pm$ 0.087	&	-0.635 $\pm$ 0.027	&	6.330 $\pm$ 0.121	&	-1.548 $\pm$ 0.084	&	--- $\pm$ ---	&	--- $\pm$ ---	\\
174 (K)	&	15360 $\pm$ 1362	&	100 $\pm$ ---	&	10.942 $\pm$ 0.065	&	5.922 $\pm$ 0.274	&	-1.842 $\pm$ 0.254	&	7.763 $\pm$ 0.106	&	7.049 $\pm$ 0.125	&	-0.714 $\pm$ 0.039	&	5.517 $\pm$ 0.081	&	-2.246 $\pm$ 0.050	&	5.497 $\pm$ 0.132	&	-2.266 $\pm$ 0.109	\\
194 (K)	&	19760 $\pm$ 2267	&	54.1 $\pm$ 37.0	&	10.950 $\pm$ 0.061	&	6.405 $\pm$ 0.038	&	-1.103 $\pm$ 0.032	&	7.508 $\pm$ 0.052	&	6.805 $\pm$ 0.078	&	-0.703 $\pm$ 0.046	&	5.722 $\pm$ 0.028	&	-1.785 $\pm$ 0.029	&	5.228 $\pm$ 0.091	&	-2.281 $\pm$ 0.089	\\
218 (K)	&	12800 $\pm$ 435	&	100 $\pm$ ---	&	10.889 $\pm$ 0.018	&	6.474 $\pm$ 0.042	&	-1.587 $\pm$ 0.025	&	8.061 $\pm$ 0.043	&	7.401 $\pm$ 0.051	&	-0.662 $\pm$ 0.015	&	6.340 $\pm$ 0.100	&	-1.721 $\pm$ 0.086	&	5.623 $\pm$ 0.056	&	-2.439 $\pm$ 0.045	\\
303 (K)	&	18140 $\pm$ 2715	&	83.2 $\pm$ 362	&	10.874 $\pm$ 0.076	&	5.486 $\pm$ 0.143	&	-1.588 $\pm$ 0.113	&	7.076 $\pm$ 0.123	&	6.246 $\pm$ 0.160	&	-0.830 $\pm$ 0.075	&	4.649 $\pm$ 0.103	&	-2.425 $\pm$ 0.078	&	--- $\pm$ ---	&	--- $\pm$ ---	
\enddata
\end{deluxetable*}

\section{Discussion} % Section 4.
\label{sec:discussion}

\indent With the abundance characteristics of our sample of \HA\ Dots established in Section \ref{sec:analysis}, we now examine the importance of these metal-poor systems to our understanding of star formation and chemical enrichment at low luminosities and stellar masses.
\\
\indent Previous studies of magnitude-limited star-forming galaxy samples generally favor detection of bright systems (i.e., the Malmquist effect).
Comprehensive understanding of the nature of such galaxies, and in particular their connection with the overall distribution of galaxies, is therefore skewed toward the properties exhibited by these more conspicuous systems (see \citealp{bib:Hirschauer2018} for a more in-depth discussion).
\\
\indent Cosmological simulations have long predicted that small, low-luminosity systems should constitute a large fraction of the number density of galaxies (e.g., \citealp{bib:Mateo1998}).
The observed number density of dwarf galaxies in the solar neighborhood and nearby clusters, however, is far less than predicted.
This mismatch between simulations and observations is known as the ``Missing Satellites Problem" and has large implications regarding the accuracy of predicting the evolution of matter distributions (e.g., \citealp{bib:Moore1999, bib:Klypin1999, bib:Koposov2008, bib:BullockBoylan-Kolchin2017}).
To the contrary, we expect the actual number density of low-luminosity systems to be quite high.
\\
\indent Furthermore, these galaxies are thought to have contributed significantly to the overall star formation and chemical and dust enrichment history of the universe, particularly at early epochs near Cosmic Noon ($z$ $\sim$ 1.5-2; \citealp{bib:MadauDickinson2014}).
Detailed study of local analogs to higher-redshift galaxies and an understanding of their characteristics is therefore invaluable to the understanding of one of the major producers of stars, metals, and dust in the formative years of the universe.

\subsection{Improving the Census of Low-Luminosity and Low-Mass Systems} % Section 4.1.
\label{sec:census}

\indent The serendipitous discovery of \HA\ Dots within the AHA survey has provided an exciting opportunity to help populate the low-luminosity end of the overall distribution of star-forming galaxies, and to build a link between studies focusing on these dwarf systems with larger-scale abundance studies of more luminous and metal-rich samples.
Because of their low luminosities, such dwarf systems are unlikely to be targeted by large-scale surveys.
Projects such as SDSS \emph{do} encompass low-luminosity dwarf galaxies, including \HA\ Dots, but because of their general unremarkableness, often receive little attention for campaigns of follow-up study.
For example, of the twenty-six \HA\ Dots with \Te\ abundances in the current study, only four ($\sim$15.4\%) were observed as part of the Sloan spectroscopic component \citep{bib:Ahn2012}.
The lowest-luminosity star-forming sources are, understandably, comparatively under-represented in magnitude-limited studies of chemical abundances in contrast to more luminous sources.
\\
\indent We present the metallicity information for the compact, low-luminosity, low-mass, dwarf star-forming galaxies of this study in Table \ref{tab:abundances_all} (see \S \ref{sec:analysis} for details).
These dwarf galaxies span a luminosity range of $M_{R}$ = --17.61 to --11.02, with an average $R$-band absolute magnitude of --15.70.
Using the photometric conversion of \citet{bib:Cook2014}, this translates to $M_{B}$ = --17.04 to --10.75, with an average $B$-band absolute magnitude of --14.96 (see \S \ref{sec:LZR} for details).
With determinations of stellar masses by SED model fitting as detailed in \S\ref{sec:MassAndSFR}, and illustrated in Figure \ref{fig:sSFR}, we additionally find that the \HA\ Dots sample is lower mass on average than both the star-forming galaxies from the KISS survey and green pea systems, though slightly more massive on average than the blueberries.
Stellar masses of our sample were determined to span between log $M_{*}$ = 5.55 and 8.67, with an average value of log $M_{*}$ = 7.71.
\\
\indent \HA\ Dot ranges for both $B$-band luminosity and stellar mass are roughly consistent with those demonstrated by the ``Combined Select" sample of dwarf star-forming galaxies of the Local Volume Legacy (LVL) project from \citet{bib:Berg2012}, a carefully-curated collection of thirty-eight objects acquired in part from the literature.
The luminosity range demonstrated by these LVL galaxies is $M_{B}$ = --18.02 to --10.91, with an average of $M_{B}$ = --14.40.
Similarly, stellar mass estimates of the LVL galaxies span between log $M_{*}$ = 5.9 and 9.3, with an average value of log $M_{*}$ = 7.6.
Our twenty-six \HA\ Dots provide robust abundance measurements over an equivalent range of luminosity and stellar mass, and therefore represents an important significant addition to the census of sources in these regimes.
These constitute a noteworthy increase in the number of such systems known in the local universe, and which occupy the sparsely-populated extreme faint end of the star-forming galaxy luminosity distribution.
% with robust abundances

\subsection{Relative Abundances} % Section 4.2.
\label{sec:relative_abundances}

\begin{figure*} % Figure 5
\figurenum{5}
\plotone{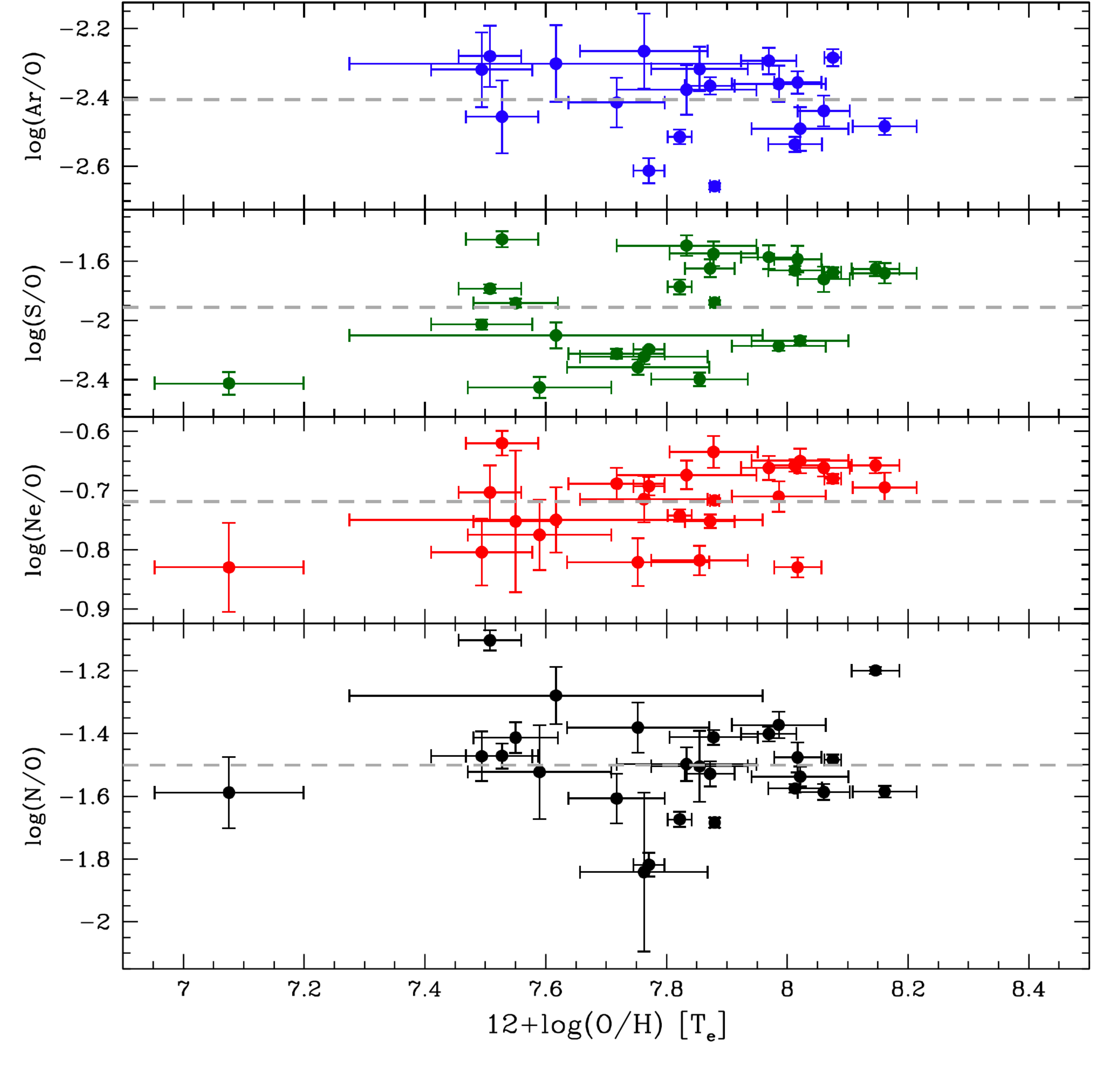}
\caption{
Elemental abundances for nitrogen, neon, sulfur, and argon as a function of oxygen abundance.
Dashed lines represent the average value for each element across the \HA\ Dots sample.
Consistent with expectation for such a low-metallicity sample, the alpha elements (neon, sulfur, and argon) remain roughly constant with increasing oxygen abundance.
The nitrogen abundance similarly stays roughly constant, reflecting primary nitrogen production alone, which is expected at low metallicity.
}
\label{fig:alpha_elements}
\end{figure*}

\indent Abundance ratios of nitrogen, neon, sulfur, and argon of the \HA\ Dots sample are presented in Figure \ref{fig:alpha_elements}, plotted as a function of oxygen abundance.
We note that some abundance values exhibit large uncertainties due to the propagated large uncertainties of the relevant \Te\ values.
Mean values for each element across our \HA\ Dots sample (log(N/O) = --1.50 $\pm$ 0.06, log(Ne/O) = --0.72 $\pm$ 0.03, log(S/O) = --1.91 $\pm$ 0.05, and log(Ar/O) = --2.41 $\pm$ 0.05) are illustrated as dashed lines.
These averages for the alpha elements neon and sulfur are roughly consistent with dwarf irregular (e.g., \citealp{bib:vanZee1997, bib:vanZeeHaynes2006, bib:Skillman2013, bib:Hirschauer2016}) and starbursting dwarf galaxies (e.g., \citealp{bib:Thuan1995, bib:Izotov1997a, bib:IzotovThuan1998}) in the literature.
Our average argon abundance, however, appears to be somewhat lower.
This may be as a consequence of \ArIII$\lambda$7136 presenting in a very noisy region of the spectrum.
The low value of the average argon abundance and the large dispersion in the argon and sulfur abundances may be due in part to choice of ICFs.
In addition, the uncertainties in the sulfur abundance values may be inflated due to the necessity of using the intrinsically-weak, high-ionization \SIII$\lambda$6312 line as our only measurement of the S$^{++}$ ionic abundance.
%some abundance values are high due to the necessity of using intrinsically-weak high-ionization emission lines, including \SIII\ for sulfur and \ArIV\ for argon.
The value of average \HA\ Dot log(N/O) ratio is greater than both the plateau at --1.60 defined for XMPs by \citet{bib:IzotovThuan1999} and what is found in \HII\ regions of very low metallicity (e.g., \citealp{bib:Thuan1995, bib:Izotov1997a, bib:IzotovThuan1998}), but is comparable with that of nearby, metal-poor, dwarf irregular galaxies (e.g., \citealp{bib:Skillman2003, bib:VilchezIglesias-Paramo2003, bib:vanZeeHaynes2006, bib:Berg2012, bib:Skillman2013, bib:Hirschauer2016}).
\\
\indent Overall the alpha elements exhibit no trend as the oxygen abundance increases, remaining approximately constant across the parameter space, a behavior that is consistent with expectation for metal-poor sources (e.g., \citealp{bib:vanZee1998, bib:IzotovThuan1999, bib:Berg2012, bib:Izotov2012, bib:Skillman2013}).
The nitrogen-to-oxygen ratio similarly remains constant with increasing oxygen abundance, an indication that these \HA\ Dots are experiencing predominantly primary nitrogen production (e.g., \citealp{bib:Vila-CostasEdmunds1993, bib:vanZee1998, bib:Henry2000}).
At moderate to high metallicities (\abun\ $>$ 7.9; \citealp{bib:Torres-Peimbert1989, bib:KobulnickySkillman1996, bib:KobulnickySkillman1998}), the contribution of secondary nitrogen production becomes manifest as a linear increase of log(N/O) with log(O/H).
Metallicity values for our sample remain at or below this threshold, and thus such an upturn in the plot is not seen.

\subsection{The Luminosity-Metallicity Relation with Low-Luminosity Sources} % Section 4.3.
\label{sec:LZR}

\begin{figure*} % Figure 6
\figurenum{6}
\plotone{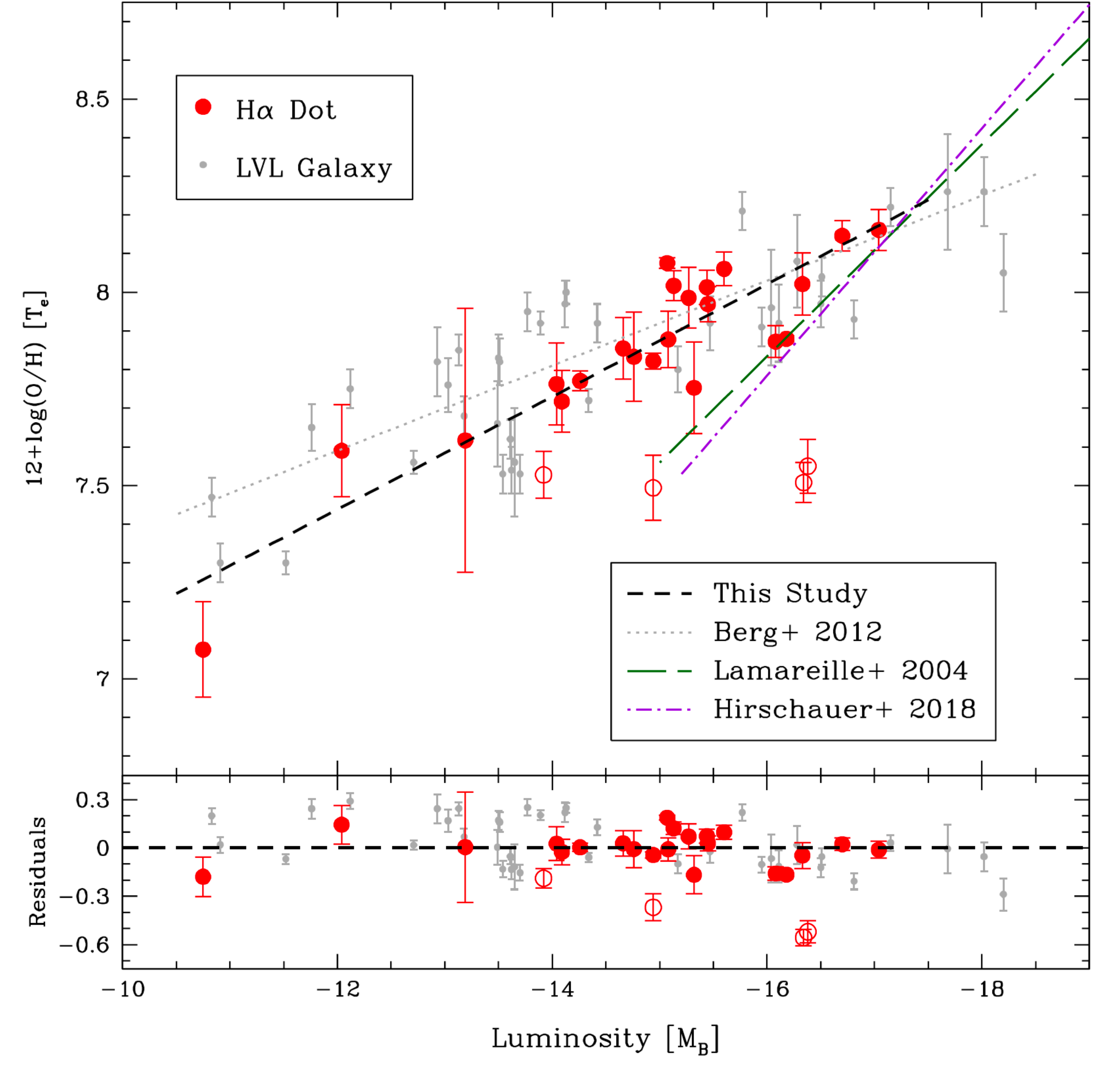}
\caption{
\emph{B}-band luminosity-metallicity relation (\LZR) plot for \HA\ Dots (red points) and LVL ``Combined Select" galaxies (gray points; \citealp{bib:Berg2012}), all with \Te-method oxygen abundances.
% robust,
A linear fit to the \HA\ Dots data is included as a black dashed line, with residuals to the fit in the lower portion of the plot.
%Open red circles represent \HA\ Dots with suspected influence by pristine infall; we have opted to remove them from the fit.
Open red circles represent \HA\ Dots with large \Te\ uncertainties and suspected over-inflated \Te\ values, manifest as under-estimated metallicities; we have opted to remove them from the fit (see \S\ref{sec:LZR} for more details).
The ``Combined Select" LVL dwarf galaxy sample \LZR\ fit from \citet{bib:Berg2012} is also included as a dotted gray line.
\LZR\ fits to star-forming galaxy samples spanning a full luminosity range are presented as well;
\citet{bib:Lamareille2004} as a long-dashed green line and \citet{bib:Hirschauer2018} as a dot-dashed violet line.
\HA\ Dots are found to inhabit this lower-luminosity end of the star-forming galaxy distribution.
The shallow slope suggests that, at the lowest luminosities, such systems do not become arbitrarily metal poor.
Enrichment to a minimum value therefore occurs rapidly over the course of galaxy evolution.
}
\label{fig:LZR}
\end{figure*}

\begin{deluxetable}{lccc} % Table 4.
%\rotate
\tablenum{4}
\tabletypesize{\scriptsize}
\tablewidth{0pt}
\tablecaption{Polynomial Coefficients for Functional Forms of Luminosity-Metallicity Relation Fits Available in the Literature}
\tablehead{\colhead{\LZ\ Study}&\colhead{$A$}&\colhead{$B$}&\colhead{RMS}}
\startdata
\noindent Full Luminosity Range \\
\hline
\noindent \citet{bib:Lamareille2004} & 3.45 $\pm$ 0.09 & --0.274 $\pm$ 0.005 & 0.27 \\
\noindent \citet{bib:Hirschauer2018} & & & \\
\noindent $\rightarrow$ [KISSR SEL] & 2.664 $\pm$ 0.170 & --0.320 $\pm$ 0.009 & 0.280 \\
\hline																							
\noindent Low-Luminosity Dwarfs Only \\
\hline																							
\noindent \citet{bib:Berg2012} & & & \\
\noindent $\rightarrow$ [``Combined Select"] & 6.27 $\pm$ 0.21 & --0.11 $\pm$ 0.01 & 0.15 \\
\noindent \citet{bib:Hirschauer2018} & & & \\
\noindent $\rightarrow$ [KISSR \Te] & 6.543 $\pm$ 0.116 & --0.084 $\pm$ 0.007 & 0.216 \\
\noindent This work & 5.692 $\pm$ 0.232 & --0.146 $\pm$ 0.016 & 0.099 \\
\enddata
\label{tab:LZpolycfs}
\tablecomments{Polynomial coefficients are presented in the form of \abun~=~$A$~+~$Bx$, where $x$~=~$M_{B}$.}
\end{deluxetable}

\indent Scaling relationships between a galaxy's metallicity and stellar content can provide clues to better understand chemical enrichment processes and star-formation history.
The production of metals reflects nucleosnythetic processing as a consequence of successive generations of stars within a given system, and the luminosity of a galaxy directly traces stellar content by measured starlight.
Consequently, we find that a galaxy's brightness scales positively with its level of enrichment.
This correlation between luminosity and metallicity (the \LZ\ relation, or \LZR) is demonstrated across a wide parameter space along both dimensions (e.g., \citealp{bib:Skillman1989, bib:Vilchez1995, bib:Lamareille2004, bib:Tremonti2004, bib:Salzer2005, bib:Hirschauer2018}).
Some complicating phenomena must be taken into consideration, however, before a complete picture is established.
These include enriched gas expulsion by supernovae (made more efficient in dwarf galaxies by weak gravitational potential wells), galaxy interactions and mergers inducing new starbursts, luminosity enhancements and abundance dilution created by infalling pristine hydrogen gas, and the effects of stellar feedback.
\\
\indent As low-luminosity sources are often under-represented in flux- or magnitude-limited surveys of star-forming galaxies, regressions made to an \LZR\ can be biased toward brighter, more massive systems.
Efforts to establish fits to such an \LZR\ will inevitably produce models which more accurately reflect the behaviors of these large- and medium-sized galaxies, but may fail in the regime of dwarfs.
In particular, the resulting steep linear fits imply that systems in the lowest range of the luminosity distribution should have metallicities well below what observations currently measure and, generally, that an arbitrarily low metallicity is possible.
It has been observed that for samples consisting of exclusively low-luminosity galaxies, slopes of \LZR\ fits are shallower than for samples which span the full range of star-forming galaxy luminosities (e.g., \citealp{bib:Skillman1989, bib:RicherMcCall1995, bib:Lee2004, bib:vanZeeHaynes2006, bib:vanZee2006, bib:Guseva2009, bib:Berg2012, bib:Haurberg2015, bib:Hirschauer2018}).
\LZ\ studies focusing on low-luminosity dwarf systems are therefore more likely to accurately represent the star formation and chemical enrichment behavior specific to this regime.
\\
\indent We present an \LZR\ plot of the twenty-six \HA\ Dots from our sample in $B$-band with \Te-method abundances (red dots) in Figure \ref{fig:LZR}, alongside the ``Combined Select" sample of LVL galaxies (gray dots) from \citet{bib:Berg2012}.
% robust
Overall, the \HA\ Dots inhabit only the lower-luminosity end of the overall distribution of sources, roughly comparable to the LVL sample.
The characteristics of the sample selection, in fact, preclude inclusion of brighter systems:\ such a galaxy would already have been identified by previous surveys!
We therefore expect to only include faint sources, making \HA\ Dots an ideal sample for study of the low-luminosity behavior of the \LZR.
\\
\indent A linear least squares fit to our \HA\ Dots sample is included in Figure \ref{fig:LZR} as a dashed black line, with the small, lower plot representing the residuals to the fit.
It takes the form,
\[
12+\text{log(O/H)} = 5.692(\pm0.232) - 0.146(\pm0.016) \times\ M_{B},
\]
with an \textsc{rms} scatter in \HA\ Dot abundance of $\sigma$ = 0.099.
Open red circles represent \HA\ Dots with \Te\ values possessing particularly large uncertainties, which have been subsequently excluded from the fit.
We believe that in the spectra of these sources (\HA\ Dots 20, 34, 127, and 194), the weak, temperature-sensitive \OIII$\lambda$4363 auroral line fluxes have been artificially inflated, leading to an over-estimation of the electron gas temperature and thus an under-estimation of the computed direct-method metallicity, pushing them downward off of the \LZR\ fit line.
%Substantial disagreements between the direct-method and SEL abundances for these four objects alone suggests that they are \emph{metallicity} outliers, rather than simply temperature outliers.
Removing these four outliers from the \LZR\ fit is further justified when we compare their direct-method abundances with their metallicities estimated using SEL relations.
The direct-method values for these four galaxies are substantially lower than their SEL abundances.
We find no evidence that these objects not constitute a sub-set of extreme star-forming galaxies experiencing skewed abundances (e.g., density-bounded systems with escaping Ly$\alpha$ photons which invalidate the ICFs).
%In addition, we note that, at such high values of \Te, even fairly large temperature uncertainties are manifest as moderate uncertainties in the associated abundance value, leading to relatively small error bars.}
% add another sentence here that says that effectively, for these four cases, the SEL abundances are substantially different from T_e abundances, which isn't true for the rest of the sample.
% have to present evidence for *why* we think that these are dubious.
\\
\indent The dotted gray line is the \LZ\ fit to the ``Combined Select" LVL galaxy sample from \citet{bib:Berg2012}, and is included for comparison as an additional example of dwarf star-forming systems with \Te-method abundance measurements.
% robust
In addition, \LZR\ fits to samples spanning the full range of star-forming galaxy luminosities are illustrated in Figure \ref{fig:LZR}.
These include that of \citet{bib:Lamareille2004}, based on the 2dF Galaxy Redshift Survey (2dFGRS), and \citet{bib:Hirschauer2018}, based on the KISS survey.
They are illustrated as a long-dashed green line and a dot-dashed violet line, respectively.
\\
\indent A summary of some \LZR\ fits from the literature is presented in Table \ref{tab:LZpolycfs}, including studies which examined the full range of star-forming galaxy luminosities and those which focused on low-luminosity dwarf systems only.
Because obtaining direct-method abundance measurements is limited to only fairly metal-poor systems, studies covering the full luminosity range rely on SEL empirical relations calibrated to sources of known abundance (e.g., \citealp{bib:Tremonti2004}).
% bib:KewleyEllison2008
In order to directly compare these large-scale surveys with smaller samples of only low-luminosity systems possessing \Te-method abundances (including the \HA\ Dots), we must ensure that the metallicity scales are comparable \citep{bib:KewleyEllison2008}.
Metallicity relation fits made to SEL abundance estimates of the \HA\ Dots sample were produced utilizing McGaugh model abundance grids \citep{bib:McGaugh1991} as well as the O3N2 method calibration presented in \citet{bib:Hirschauer2018}.
We found no appreciable difference to our \LZR\ fits made utilizing the direct-method metallicity values as compared to those determined with SEL methods.
Due to its exhibiting smaller uncertainty, we have opted to retain the \LZR\ fit made using \Te-method abundances for the \HA\ Dots sample.
\\
\indent From Figure \ref{fig:LZR} we see that our \LZR\ relation fit traces the \HA\ Dots and LVL ``Combined Select" dwarf galaxy sample well, and is only slightly steeper than the fit presented by \citet{bib:Berg2012}.
\LZR\ fits to samples spanning the full range of star-forming galaxy luminosities, however, are found to be much steeper.
Both such fits from \citet{bib:Lamareille2004} and \citet{bib:Hirschauer2018} would, if extrapolated to lower luminosities, fail to accurately represent the dwarf galaxy samples presented in Figure \ref{fig:LZR}.
This dependency on the range of sampled luminosity provides insight to the evolutionary behavior of dwarf star-forming systems.
Consistent with the findings of \citet{bib:Blanc2019}, such a flattening in the \LZR\ slope is consistent with the idea that heavy element enrichment of even the smallest galaxies will reach a minimum value rapidly in their evolutionary history \citep{bib:KunthSargent1983}.
\\
\indent The necessity for robust abundance measurements to inform \LZR\ fits of low-luminosity galaxies alone is therefore clear:
In order to properly characterize the astrophysics associated with chemical enrichment and evolution of these dwarf systems, we require high-quality spectral data from star-forming galaxies at the lowest luminosities, such as the \HA\ Dots sample.
%\\
%%
%\indent Finally, we make note here that the \HA\ Dot with the highest residual to the \LZR\ fit is \HA\ Dot 303 ($M_{B}$ = --10.75).
%This system is, additionally, that which possesses the lowest oxygen abundances in our sample, with \abun\ = 7.076 $\pm$ 0.123.
%The astrophysical mechanism behind the offset of \HA\ Dot 303 from the \LZR\ fit is expected to arise from a luminosity enhancement by recent starburst (moving right in the \emph{x}-direction; see \citealp{bib:Hirschauer2016} and \citealp{bib:McQuinn2020} for more details).
% in part from an influx of pristine gas, which both dilutes the metallicity (moving down in the \emph{y}-direction) and drives
%For \HA\ Dot 303 to be positioned where it is now on the \LZR\ diagram, pushed significantly off the fit line, it must have originated as an \emph{exceptionally} low-luminosity, metal-poor dwarf galaxy.

\subsection{The Stellar Mass-Metallicity Relation with Low-Mass Sources} % Section 4.4.
\label{sec:MZR}

\begin{figure*} % Figure 7
\figurenum{7}
\plotone{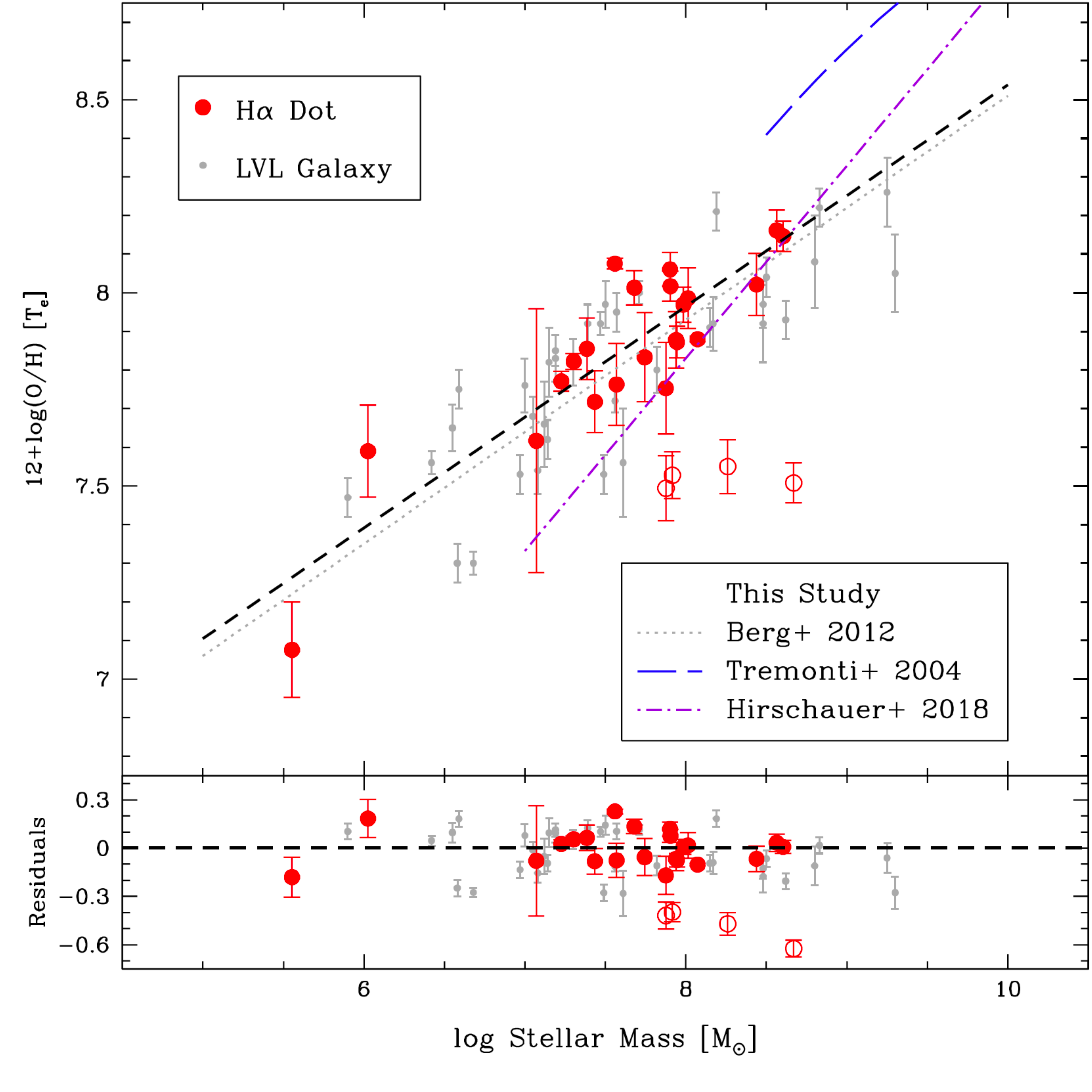}
\caption{
Stellar mass-metallicity relation {\MZR} plot of \HA\ Dots (red points) and LVL ``Combined Select" galaxies (gray points; \citealp{bib:Berg2012}), all with $T_{e}$-method oxygen abundances.
% robust, 
%Open red circles represent \HA\ Dots with suspected influence by pristine infall; we have opted to remove them from the fit.
Open red circles represent \HA\ Dots with large \Te\ uncertainties and suspected over-inflated \Te\ values, manifest as under-estimated metallicities; we have opted to remove them from the fit (see \S\ref{sec:LZR} for more details).
Estimates of \HA\ Dot stellar masses were determined by SED model fit analyses.
A linear fit to the \HA\ Dots data is included as a black dashed line, with residuals to the fit in the lower portion of the plot.
It is nearly identical to the ``Combined Select" LVL dwarf galaxy sample \MZR\ fit from \citet{bib:Berg2012}, which is also included as a dotted gray line.
\MZR\ fits to star-forming galaxy samples spanning a wider mass range are presented as well;
\citet{bib:Tremonti2004} as a long-dashed blue line and \citet{bib:Hirschauer2018} as a dot-dashed violet line.
Like the \LZR\ presented in Figure \ref{fig:LZR}, we find that \HA\ Dots successfully populate the lower-mass range of the distribution of star-forming galaxies, with a fit slope that is shallower than that constructed from wider-spanning samples.
}
\label{fig:MZR}
\end{figure*}

\indent Stellar mass is a more direct representation of stellar content, and therefore acts as an important alternative for study of metallicity scaling relations.
Like the \LZR, the stellar mass-metallicity relation (\MZR) holds across a large range of parameter space, and exhibits a relatively small amount of scatter (e.g., \citealp{bib:Tremonti2004, bib:KewleyEllison2008, bib:Zahid2011, bib:Zahid2012, bib:Berg2012, bib:AndrewsMartini2013, bib:Haurberg2015, bib:Hirschauer2018, bib:Indahl2021}).
Unlike luminosity, however, determination of stellar mass is not a direct observable and is consequently more difficult to ascertain reliably.
The stellar mass of a given system is typically computed via SED-fitting routines which utilize multiwavelength photometric data, and account for effects such as dust absorption and star formation in a manner that adoption of a single mass-to-light ($M_{*}$/$L$) ratio conversion cannot.
\\
\indent The \MZR\ for our \HA\ Dots sample is presented as Figure \ref{fig:MZR}.
Here we see our sample of twenty-six sources (red dots), plotted with the LVL ``Combined Select" galaxies of \citet{bib:Berg2012}.
The black dashed line constitutes a linear least squares fit to the \HA\ Dots data, taking the form,
\[
12+\text{log(O/H)} = 5.671(\pm0.264) + 0.287(\pm0.035) \times\ \text{log M$_{*}$},
\]
with an \textsc{rms} scatter in \HA\ Dot abundance of $\sigma$ = 0.109.
Again, open red circles represent \HA\ Dots with very large \Te\ uncertainties and suspected inflated temperatures, resulting in under-estimates of the direct-method metallicities (\HA\ Dots 20, 34, 127, and 194); we have opted to remove them from the fit (see \S\ref{sec:LZR} for more details).
The gray dotted line is a fit to the \citet{bib:Berg2012} LVL sample and is nearly coincident with that of the \HA\ Dots.
Included as well are two \MZ\ fits made to samples spanning much wider ranges of stellar mass.
The blue long-dashed line shows the low-mass end of the \citet{bib:Tremonti2004} relation made to SDSS, while the violet dot-dashed line is the fit to the full KISS sample from \citet{bib:Hirschauer2018}.
All polynomial coefficients for the fit lines are summarized in Table \ref{tab:MZpolycfs}.
\\
\indent A similar trend as is seen with the \LZR\ of Figure \ref{fig:LZR} is clearly evident.
Metallicity scaling relations made to samples of lower stellar mass are found to be shallower than those constructed to wider-scale collections of galaxies.
%The empirical-relation abundances fit to the \HA\ Dots was thus rejected in favor of that utilizing \Te-methods, which presents smaller uncertainty.
This suggests that chemical enrichment in low-mass star-forming systems proceeds differently than what is expected based on studies made to larger samples, in agreement with the discussion of \S\ref{sec:LZR}.
Our results are consistent with those from the study of \citet{bib:Blanc2019}, which used a combination of observational data and hydrodynamical simulations.
Again, \MZR\ fits made to SEL-method \HA\ Dot abundances were produced to compare the large-scale sample with that at low stellar masses only, and were found to show no appreciable difference to that made utilizing direct-method metallicities.
\\
\indent We note that the most metal-poor object in our sample, \HA\ Dot 303 (\abun\ = 7.08 $\pm$ 0.12; log M$_{*}$ = 5.55 $M_{\odot}$), presents a modest offset from the \MZR\ fit line.
%With a stellar mass estimate of log M$_{*}$ = 5.47 $M_{\odot}$, this system lies offset just below  the \MZR\ fit line.
While \HA\ Dot 303's extremely low metallicity is likely a consequence of inefficient star formation and metal loss via stellar feedback-dominated galactic winds, the small displacement evident here and in Figure \ref{fig:LZR} is likely due to high levels of recent star formation induced by a minor interaction with the nearby companion dwarf irregular UGC 5186.
Neutral hydrogen observations show tentative evidence for a gas bridge connecting the two systems, providing the means for a modest infall \citep{bib:McQuinn2020}.

\begin{deluxetable}{lcccc} % Table 5.
%\rotate
\tablenum{5}
\tabletypesize{\scriptsize}
\tablewidth{0pt}
\tablecaption{Polynomial Coefficients for Functional Forms of Stellar Mass-Metallicity Relation Fits Available in the Literature}
\tablehead{\colhead{\MZ\ Study}&\colhead{$A$}&\colhead{$B$}&\colhead{$C$}&\colhead{RMS}}
\startdata
\noindent Full Stellar Mass Range \\
\hline
\noindent \citet{bib:Tremonti2004} & --1.492 & 1.847 & --0.08026 & 0.27 \\
\noindent \citet{bib:Hirschauer2018} & & & \\
\noindent $\rightarrow$ [KISSR SEL] & 3.838 $\pm$ 0.102 & 0.499 $\pm$ 0.007 & \ldots & 0.182 \\
\hline																							
\noindent Low-Mass Dwarfs Only \\
\hline																							
\noindent \citet{bib:Berg2012} & & & \\
\noindent $\rightarrow$ [``Combined Select"] & 5.61 $\pm$ 0.24 & 0.29 $\pm$ 0.03 & \ldots & 0.15 \\
\noindent \citet{bib:Hirschauer2018} & & & \\
\noindent $\rightarrow$ [KISSR \Te] & 5.990 $\pm$ 0.118 & 0.236 $\pm$ 0.014 & \ldots & 0.198 \\
\noindent This work & 5.671 $\pm$ 0.264 & 0.287 $\pm$ 0.035 & \ldots & 0.109 \\
\enddata
\label{tab:MZpolycfs}
\tablecomments{Polynomial coefficients are presented in the form of \abun~=~$A$~+~$Bx$~+~$Cx^{2}$, where $x$~=~log M$_{*}$.}
\end{deluxetable}

\subsection{The Most Metal-Poor Star-Forming Galaxies} % Section 4.5.
\label{sec:XMPs}

\indent The nearby dwarf star-forming galaxy contingent of the \HA\ Dots sample comprise some of the lowest-abundance systems known.
With selection criteria emphasizing environmental isolation, an important factor for metal-poor systems (e.g., \citealp{bib:McQuinn2015b}), coupled with conspicuous \HA\ emission, we find that these sources' observational characteristics enjoy a significant overlap in parameter space with the realm of XMPs, suggesting that yet more such specimens exist within the \HA\ Dots sample.
These extreme systems are defined as having an oxygen abundance of \abun\ $\leq$ 7.35, or approximately 5\% $Z_{\odot}$ \citep{bib:McQuinn2020}.
But XMP galaxies are \emph{exceptionally} uncommon, with only a scant handful of examples known in the local universe (see \citealp{bib:McQuinn2020} for a recent review of this topic).
While preliminary spectral observations have been completed for all of the sources presented in the first two survey papers \citep{bib:Kellar2012, bib:Salzer2020}, nearly 180 additional \HA\ Dots have yet to be analyzed.
Subsequent detailed spectroscopy may reveal more XMPs within the sample.
\\
\indent Nearby XMPs represent our best accessible analogs to the kinds of star-forming systems which populated the universe during Cosmic Noon, when a substantial amount of the universe's star formation and chemical enrichment took place.
Detailed abundance analyses of multiple such galaxies gives us our most robust opportunity from which to base models of the early universe, where direct observations are precluded.
As described in \S \ref{sec:LZR}, the lowest-abundance systems are expected to be quite faint, making direct detection difficult in the first place.
\\
\indent While some recent success has been achieved in discovering new XMP galaxies within directed photometric surveys (e.g., \citealp{bib:Guseva2017, bib:Hsyu2017, bib:Izotov2018, bib:Hsyu2018, bib:SenchynaStark2019, bib:Kojima2020, bib:Izotov2021}), these are typically accomplished by mining existing data sets such as SDSS for specific photometric qualities.
In contrast, indirect search methods such as the \HA\ Dots' serendipitous detections in the AHA imaging survey, or the new star formation potential of neutral gas reservoirs as identified by the ALFALFA and SHIELD surveys (leading to the discovery of both Leo P and Leoncino), are an exciting prospect for future endeavors.
\\
\indent Perhaps the ongoing follow-up spectroscopy campaign for the remaining \HA\ Dots identified from the AHA project WIYN 0.9m images, currently being undertaken with the HET and the updated Low Resolution Spectrometer (LRS2) in queue mode, will reveal additional XMP candidates.
From recently-processed AHA photometry utilizing the KPNO 2.1m telescope\footnote{The KPNO 2.1-m telescope was formerly operated by the National Optical Astronomy Observatory (NOAO), which consisted of KPNO near Tucson, Arizona, Cerro Tololo Inter-American Observatory near La Serena, Chile, and the NOAO Gemini Science Center.
NOAO was operated by the Association of Universities for Research in Astronomy (AURA) under a cooperative agreement with the National Science Foundation},
the deeper imaging of \citet{bib:Watkins2021} has produced a catalog of 454 newly-discovered compact extragalactic sources.
Based on preliminary analysis of their compact but extended morphology, roughly 50 of these are likely low-redshift, low-luminosity candidates with strong emission lines for high-quality spectral follow-up.
Through our continuing spectroscopic campaign, we anticipate the \HA\ Dots project to supply important abundance-quality data for additional newly-discovered XMP galaxies.

\section{Summary} % Section 5.
\label{sec:summary}

\indent In this paper we present high-quality optical emission-line spectral data of twenty-six low-luminosity dwarf star-forming galaxies from the \HA\ Dots survey.
These objects are quite metal poor and represent among the lowest-luminosity and stellar mass sources observable in the distribution of nearby star-forming galaxies.
With robust chemical abundances determined by the ``direct"-, or ``\Te"-method, \HA\ Dots are invaluable for characterizing both elemental abundance properties of metal-poor sources as well as the low-luminosity/low-mass portions of the luminosity-metallicity and stellar mass-metallicity scaling relations (\LZR\ and \MZR, respectively).
\\
\indent We have found that the alpha element abundances of \HA\ Dots are consistent with other metal-poor dwarf star-forming galaxies in the local universe.
Furthermore, the alpha elemental abundance values remain constant with increasing oxygen abundance, in alignment with expectation for low-abundance systems.
At the low levels of chemical enrichment measured for these sources, nitrogen abundances similarly remain constant with increasing oxygen, indicating that primary nitrogen synthesis is the principle method.
\\
\indent Additionally, metal abundance properties of \HA\ Dots at low luminosities exhibit a flattening of the slope of the \LZR.
These findings are consistent with those of the recent study of \citet{bib:Blanc2019}, and imply that, at increasingly fainter absolute magnitudes, a star-forming galaxy will not become arbitrarily metal poor.
Instead, even at the lowest luminosities, these systems self-enrich to some minimum level of enrichment early on in their evolutionary history.
Similarly, we find a flattening of the \MZR\ slope at low stellar masses, similarly supported by the results of \cite{bib:Blanc2019}.
These findings have significant implications regarding the enrichment mechanisms of the smallest star-forming galaxies commonplace at Cosmic Noon, responsible for a significant amount of the universe's stellar and chemical enrichment, and which eventually coalesced to form the massive, giant galaxies that we see today.
\\
\indent Completion of the \HA\ Dots imaging analysis using AHA photometry from the WIYN 0.9m telescope will provide a volume-limited sample of dwarf star-forming systems extending to higher redshifts than previous samples.
Future work utilizing deeper AHA project photometry from the KPNO 2.1m telescope promises to recover yet more dwarf systems and at even lower luminosities to flesh out the census of galaxies at this extreme end of the galaxy luminosity distribution.
We will also attempt to acquire additional abundance-quality spectroscopic data of \HA\ Dots with future observational programs.
%analysis of AHA photometry from the WIYN 0.9m telescope \HA\ Dots project 
%Future observing programs, including the conclusion of the \HA\ Dots project sample using WIYN 0.9m and deeper KPNO 2.1m photometric data, promise to find yet more low-luminosity star-forming dwarfs which will flesh out the census of galaxies at the very lowest luminosities.
%The spectroscopic campaign to perform additional \HA\ Dots observations will continue, utilizing the 9.2m Hobby-Eberly Telescope (HET) and updated Low Resolution Spectrometer (LRS2) to acquire abundance-quality spectra in queue mode.
%An ongoing observing campaign, utilizing the 9.2m Hobby-Eberly Telescope (HET) and updated Low Resolution Spectrometer (LRS2), will continue acquiring new abundance-quality spectra in queue mode of additional \HA\ Dots.
\\
\indent Finally, we anticipate uncovering additional XMP sources like \HA\ Dot 303 (also known as \Alecxy, the Leoncino Dwarf) from within the \HA\ Dots survey catalog.
%with properties like \HA\ Dot 303, also known as Leoncino
%(\Alecxy), from the SHIELD sample of \HI-detected galaxies.
These extreme systems push the limits of low metallicities in observed systems and may help inform models of star formation and chemical enrichment in environments analogous to the aftermath of the Big Bang, where direct observations are impossible.
Detailed study of such accessible analogues to high-redshift star-forming galaxies is thus invaluable to a more comprehensive understanding of our universe.
\\
\\
\\
\noindent \textbf{Acknowledgements:}\
The authors would like to thank the referee for the useful comments provided which helped to improve this paper.
The work presented in this paper is based in part on observations obtained at the Kitt Peak National Observatory, National Optical Astronomy Observatory, which is operated by the Association of Universities for Research in Astronomy (AURA) under cooperative agreement with the National Science Foundation.
% (NOAO Proposal ID 2015A-0408, PI:\ J.\ Salzer)
ASH acknowledges support from NASA grant NNX14AN06G.
The ALFALFA \HA\ project, on whose data the \HA\ Dots survey project is based, was carried out with the support of the National Science Foundation (NSF-AST-0823801).
This project made use of Sloan Digital Sky Survey data.
Funding for the SDSS and SDSS-II has been provided by the Alfred P.\ Sloan Foundation, the Participating Institutions, the National Science Foundation, the U.S.\ Department of Energy, the National Aeronautics and Space Administration, the Japanese Monbukagakusho, the Max Planck Society, and the Higher Education Funding Council for England.
The SDSS Web Site is \url{https://www.sdss.org/}.
The SDSS is managed by the Astrophysical Research Consortium for the Participating Institutions.
The Participating Institutions are the American Museum of Natural History, Astrophysical Institute Potsdam, University of Basel, University of Cambridge, Case Western Reserve University, University of Chicago, Drexel University, Fermilab, the Institute for Advanced Study, the Japan Participation Group, Johns Hopkins University, the Joint Institute for Nuclear Astrophysics, the Kavli Institute for Particle Astrophysics and Cosmology, the Korean Scientist Group, the Chinese Academy of Sciences (LAMOST), Los Alamos National Laboratory, the Max-Planck-Institute for Astronomy (MPIA), the Max-Planck-Institute for Astrophysics (MPA), New Mexico State University, Ohio State University, University of Pittsburgh, University of Portsmouth, Princeton University, the United States Naval Observatory, and the University of Washington.
\\
\indent \emph{Facilities}: Mayall (Richey--Chr\'{e}tien Focus Spectrograph, KOSMOS Spectrograph), WIYN:0.9m.

%%% **************************************************************************************************** %%%

%\clearpage

\bibliography{HADots}

\end{document}